\documentclass[fleqn,usenatbib]{mnras}

\usepackage{newtxtext,newtxmath}

\usepackage[T1]{fontenc}

\usepackage{graphicx}	
\usepackage{amsmath}	

\usepackage{placeins}
\usepackage{chngcntr}
\usepackage{booktabs}
\usepackage{xcolor}

\graphicspath{{Images/}{Images/Appendix/}{Images/Gamma/}{Images/Models/}{Images/Oversample/}}

\usepackage{xspace}
\newcommand{\hessj}{HESS\,J1804$-$216\xspace}
\newcommand{\snrg}{SNR\,G8.7$-$0.1\xspace}

\newcommand{\psrj}{PSR\,J1803$-$2137\xspace}

\newcommand{\fhl}{3FHL\,J1804.7$-$2144e\xspace}

\newcommand{\gray}{$\gamma$-ray\xspace}
\newcommand{\grays}{$\gamma$-rays\xspace}
\newcommand{\kms}{\,km\,s$^{-1}$\xspace}
\newcommand{\cms}{\,cm$^2$\,s$^{-1}$\xspace}
\newcommand{\vlsr}{v$_{\mathrm{lsr}}$\xspace}
\newcommand{\xb}{x^{\beta_\gamma}\xspace}
\newcommand{\kg}{k_\gamma\xspace}
\newcommand{\bg}{\beta_{\gamma}\xspace}

\newcommand{\Ep}{E_{\rm{p}}}
\newcommand{\Jp}{J_{\rm{p}}}
\newcommand{\ESN}{E_{\rm{SN}}}
\newcommand{\Epmax}{E_{\rm{p,max}}}
\newcommand{\Epmin}{E_{\rm{p,min}}}
\newcommand{\Wp}{W_{\rm{p}}}

\newcommand{\Epesc}{E_{\rm{p,esc}}}
\newcommand{\Resc}{R_{\rm{esc}}}
\newcommand{\tesc}{t_{\rm{esc}}}
\newcommand{\fbubble}{f_{\rm{bubble}}}
\newcommand{\fdif}{f_{\rm{dif}}}

\newcommand{\tsedov}{t_{\rm{sedov}}}
\newcommand{\tsnr}{t_{\rm{SNR}}}
\newcommand{\deltap}{\delta_{\rm{p}}}
\newcommand{\Rdif}{R_{\rm{dif}}}
\newcommand{\Mej}{M_{\rm{ej}}}
\newcommand{\rsnr}{R_{\rm{SNR}}}

\usepackage[capitalise,nameinlink]{cleveref}
\crefname{equation}{equation}{equations}

\usepackage[english]{babel}
\addto\extrasenglish{
    
}
\addto\extrasenglish{
    
}

\title[\hessj]{Modelling the Gamma-Ray Morphology of \hessj from Two Supernova Remnants in a Hadronic Scenario}

\author[K. Feijen et al.]{
K. Feijen$^1$\thanks{Email: kirsty.feijen@adelaide.edu.au}, S. Einecke$^1$, G. Rowell$^1$,
C. Braiding$^{2}$\thanks{Presently at: Australian Space Agency, Adelaide, Australia},
M. G. Burton$^{2,3}$,
G. F. Wong$^{4,2}$
\\
$^{1}$School of Physical Sciences, University of Adelaide, Adelaide, SA 5005, Australia \\
$^{2}$School of Physics, The University of New South Wales, Sydney, NSW 2052, Australia \\
$^{3}$Armagh Observatory and Planetarium, College Hill, Armagh BT61 9DG, UK \\
$^{4}$Western Sydney University, Locked Bag 1797, Penrith South DC, NSW 2751, Australia \\
}

\date{Accepted XXX. Received YYY; in original form ZZZ}

\pubyear{\the\year}

\begin{document}
\label{firstpage}
\pagerange{\pageref{firstpage}--\pageref{lastpage}}
\maketitle

\begin{abstract}  
\hessj is one of the brightest yet most mysterious TeV \gray sources discovered to date. Previous arc-minute scale studies of the interstellar medium (ISM) surrounding this TeV \gray source revealed \hessj is likely powered by a mature supernova remnant (SNR) or pulsar, hence its origin remains uncertain. 
In this paper, we focus on the diffusive escape of cosmic-ray protons from potential SNR accelerators. These cosmic rays interact with the ISM to produce TeV \grays. We utilise the isotropic diffusion equation solution for particles escaping from a shell, to model the energy-dependent escape and propagation of protons into the ISM.
This work is the first attempt at modelling the spatial morphology of \grays towards \hessj, using arc-minute ISM observations from both Mopra and the Southern Galactic Plane Survey.
The spectral and spatial distributions of \grays for the two nearby potential SNR counterparts, \snrg and the progenitor SNR of \psrj, are presented here. 
We vary the diffusion parameters and particle spectrum and use a grid search approach to find the best combination of model parameters.
We conclude that moderately slow diffusion is required for both candidates. The most promising candidate to be powering the TeV \grays from \hessj in a hadronic scenario is the progenitor SNR of \psrj.
\end{abstract}

\begin{keywords}
ISM: cosmic-rays -- gamma-rays: ISM -- ISM: individual objects (\hessj)
\end{keywords}

\section{INTRODUCTION}
\label{sec:intro}
The High Energy Stereoscopic System (H.E.S.S.) has a sensitivity to \grays of energy 100\,GeV to tens of TeV. H.E.S.S. has identified numerous \gray sources in the Milky Way (or `galactic sources'), however, the exact nature of over 30\% of these sources remains unknown \citep{2018_HGPS}. These sources are also possibly the sites of cosmic-ray (CR) accelerators, the population of these sources are dominated by objects within their final stages of stellar evolution.

\begin{figure}
\includegraphics[width=\columnwidth]{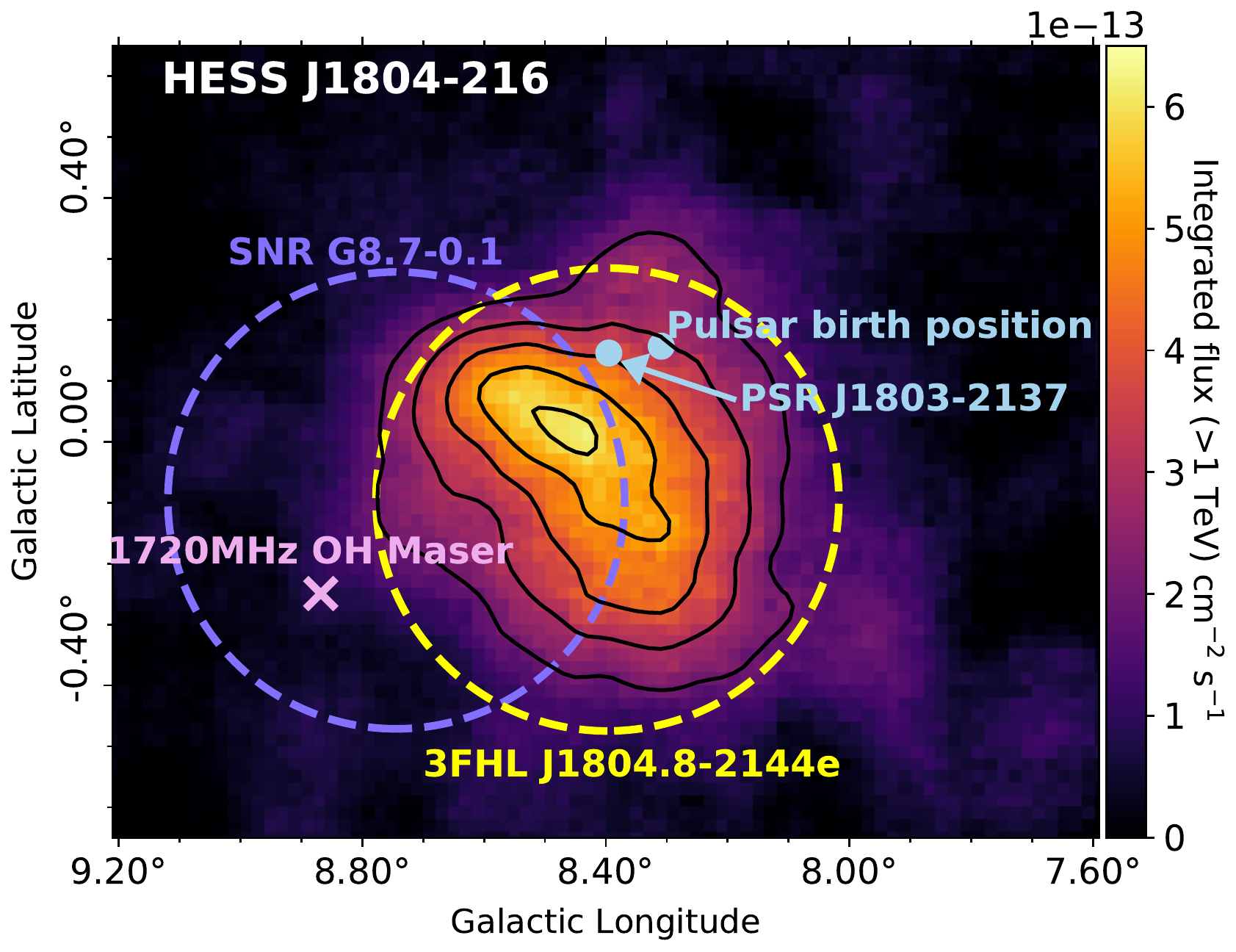}
\caption{Integrated flux map (cm$^{-2}$\,s$^{-1}$) above 1\,TeV adapted from \citet{2018_HGPS}. The dashed purple circle indicates \snrg, the purple cross indicates the 1720\,MHz OH maser and the progenitor SNR of \psrj is indicated by the blue dot. The TeV \gray emission from \citet{2018_HGPS} is shown by the solid black contours ($2\times10^{-13}$ to $6\times10^{-13}$\,cm$^{-2}$\,s$^{-1}$ levels).}
\label{fig:HESS_flux}
\end{figure}

\hessj is one of the brightest unidentified TeV \gray sources detected, with a soft spectral index of $\Gamma$=2.69. The GeV \gray source, \fhl, was detected at the same location as the TeV \gray source \hessj (see \autoref{fig:HESS_flux}). 

In a previous paper \citep{Feijen_2020}, we studied the interstellar medium (ISM) towards \hessj in detail at arc-min scales, to help determine the nature of the \gray emission. 
We investigated multiple plausible CR accelerators and concluded that a mature supernova remnant (\snrg or the progenitor SNR of \psrj) or a pulsar (\psrj) are viable accelerators of CRs to produce the TeV \gray emission in the hadronic and leptonic scenarios, respectively. 
The hadronic production of \grays involves accelerated CR protons interacting with the ISM to produce \grays through neutral pion decay \citep{Hadron_Ackermann_2013}. The leptonic scenario primarily involves TeV emission being produced by inverse-Compton upscattering by highly energetic electrons.
The TeV emission from \hessj could be produced by high-energy electrons from a pulsar wind nebulae (PWNe) powered by \psrj \citep{HESS_PWN_2018}, as supported by the high spin-down luminosity.
The modelled energy spectra of CR protons towards both \snrg and the progenitor SNR of \psrj matched with GeV and TeV observations, making it plausible for either SNR to generate the GeV and TeV \gray emission from \hessj.

SNRs are a typical candidate in accelerating CR protons at their shock front \citep{SNR_CRs_1980}. The hadronic production of \grays is investigated in this paper, assuming either \snrg or the progenitor SNR of \psrj (shown in \autoref{fig:HESS_flux}) are the plausible CR accelerators. A 1720\,MHz OH maser is present at the southern edge of \snrg at 36\kms indicating the SNR (at a velocity of 35\kms) is interacting with the ISM \citep{SNRs_Hewitt_2009}.

Models of the \gray emission from escaping CRs have previously been presented for different \gray sources, such as \citet{Casanova_2010} for SNR RX\,J1713.7$-$3946 and \citet{Mitchell_2021} for multiple SNRs. \citet{Casanova_2010} model accelerated CRs escaping SNR\,RX\,J1713.7$-$3946 in the hadronic and leptonic scenarios. Morphology maps of the \gray energy flux at 1\,TeV were obtained using the distribution of the ambient gas for different diffusion conditions. \citet{Casanova_2010} model the \gray morphology by utilising the LAB survey of HI and the Nanten survey of CO, which provides a detailed look into the distribution of \gray emission. 

This work expands on our previous paper by predicting the morphology arising from the diffusive energy-dependent escape of CR protons and interaction with the surrounding ISM in the hadronic scenario. We make use of the spherically symmetric case for the isotropic transport equation from \citet{1996_A&A}, assuming CR protons are accelerated by a single source and the energy-dependent diffusion coefficient is constant with position.
The \gray emission is modelled across all pixels of the gas column density map from the Mopra $^{12}$CO(1-0) survey and the Southern Galactic Plane Survey (SGPS) of HI for a wide range of parameters, to find the best combination of model parameters to match the observations. 
We present a first look at the 2D spatial morphology of \grays towards \hessj. This provides an important framework for understanding the region surrounding \gray sources, in particular the diffusive transport of particles from SNRs into the ISM.

\section{DATA}
\label{sec:data}
The distribution of atomic hydrogen (HI) from the SGPS\footnote{SGPS data can be found at \url{https://www.atnf.csiro.au/research/HI/sgps/fits_files.html}} \citep{SGPS_Naomi_2005} and molecular hydrogen, specifically $^{12}$CO(1-0), from the Mopra radio telescope\footnote{Published Mopra data can be found at \url{https://dataverse.harvard.edu/dataverse/harvard/}} are utilised. The Australia Telescope National Facility (ATNF) analysis software, \textsc{livedata}, \textsc{gridzilla}, and \textsc{miriad} in addition to custom \textsc{idl} routines were used to process the data from Mopra \citep{Mopra_DR0,Mopra_DR3}. Integrated emission maps were generated from the FITS cubes.

In this work, we use maps of total column density as these provide the distribution of the total target material towards \hessj. The total hydrogen column density, $N_{\rm{H}}$, is the sum of $N_{\rm{HI}}$ and 2$N_{\rm{H_2}}$, from SGPS HI observations and Mopra $^{12}$CO (regridded to the SGPS HI pixel size of $\sim$40\,arcsec).

The two plausible counterparts of interest are \snrg and the progenitor SNR of \psrj which are at distances of 4.5\,kpc \citep{SNRs_Hewitt_2009} and 3.8\,kpc \citep{PSRJ_Brisken_2006}, respectively. \snrg has an age of 15-28\,kyr \citep{SNRage_Finley_1994} and the progenitor SNR of \psrj has an age of 16\,kyr \citep{PSRJ_Brisken_2006}, assumed to be the same age as the pulsar it is attached to. 

An 1720\,MHz OH maser is indicative of interaction between SNRs and ISM clouds \citep{SNRs_Hewitt_2009}. Given the OH maser velocity of 36\,\kms, we expect \snrg to be at a similar velocity. With use of the galactic rotation curve \citep[GRC,][]{GRC_Brand_1993} and the distance to each SNR, the progenitor SNR of \psrj and \snrg are placed at a velocity of $\sim$25\,\kms and $\sim$35\,\kms, respectively. 

\begin{figure}
\includegraphics[width=\columnwidth]{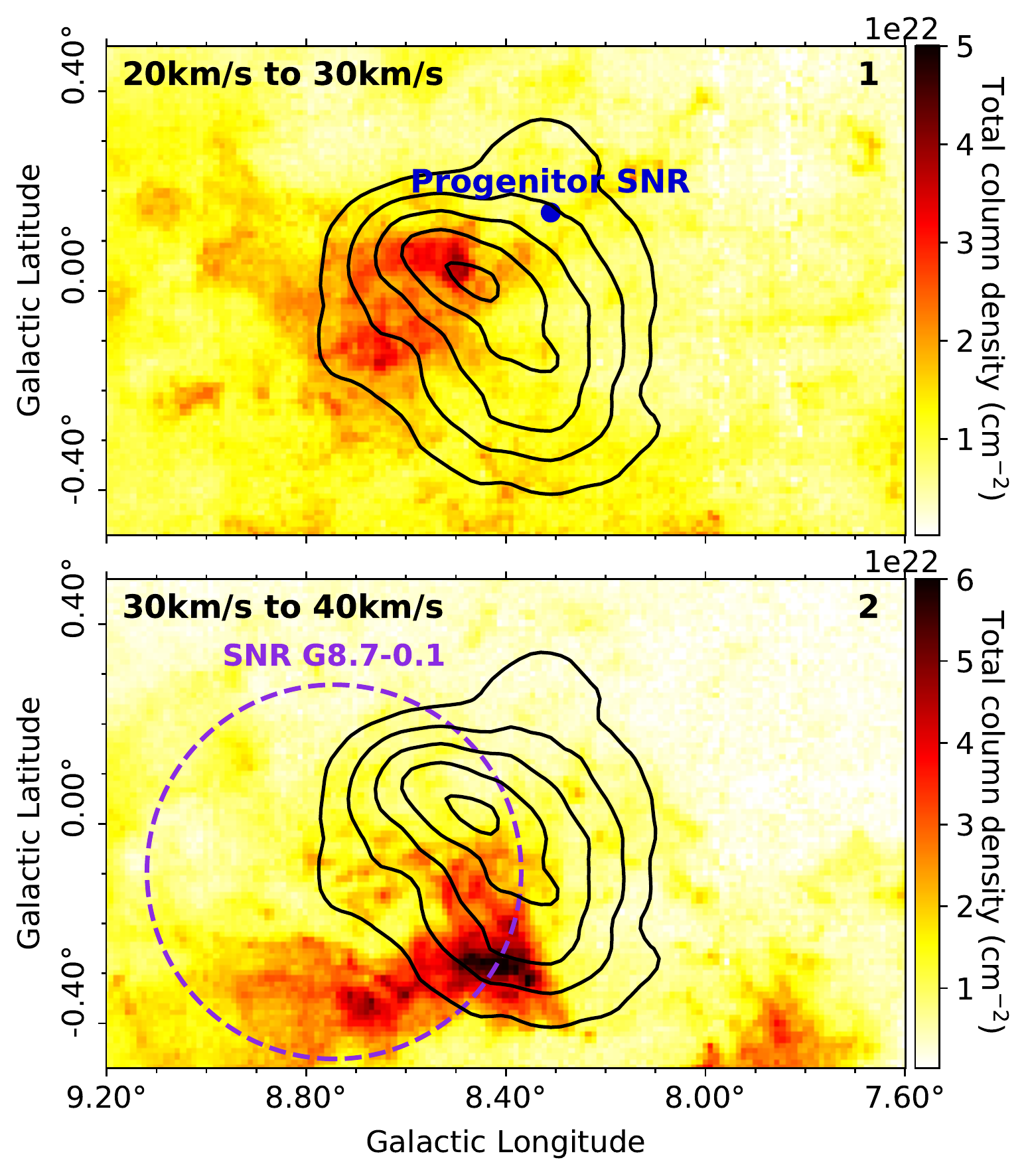}
\caption{Total column density maps, $N_{\rm{HI}}+2N_{\rm{H_2}}$, ($\rm{cm}^{-2}$) towards \hessj, for gas components 1 and 2. The progenitor SNR of \psrj is indicated by the blue dot and the dashed purple circle indicates \snrg. The TeV \gray emission from \citet{2018_HGPS} is shown by the solid black contours ($2\times10^{-13}$ to $6\times10^{-13}$\,cm$^{-2}$\,s$^{-1}$ levels).}
\label{fig:total_1_2}
\end{figure}

The velocity components are determined by taking the velocity of each counterpart as the mid-point of our range. If the velocity ranges chosen are too large, additional gas emission which is likely not connected to the source will be included. Due to local motions in the gas and the uncertainty of the GRC model, we estimate that the velocity bands should span 10\,\kms. The position-velocity plots in \autoref{fig:PVplot_CO} and \autoref{fig:PVplot_HI} show that our defined velocity regions are reasonable as they do not include too much of the gas located in the galactic arms. Currently, the SGPS HI data (\autoref{fig:PVplot_HI}) does not reveal any HI voids making it hard to narrow down these velocity ranges further. Future HI surveys, such as the GASKAP HI survey \citep{2012_GASKAP}, will have a higher resolution and be more sensitive to voids and bubbles in the HI gas.
\autoref{fig:total_1_2} shows the total column density maps derived from SGPS HI and Mopra CO for Components 1 (\vlsr=\ 20\ to\ 30\,\kms) and 2 (\vlsr=\ 30\ to\ 40\,\kms). 

The TeV \gray data used throughout this paper are from H.E.S.S.. \autoref{fig:HESS_flux} shows the \gray flux map above 1\,TeV of \hessj from the H.E.S.S Galactic Plane Survey \citep[HGPS,][]{2018_HGPS}. The HGPS \gray flux maps are available as oversampled maps which are obtained by dividing the survey region into a grid of 0.02\,$^{\circ}$, then summing all values within a circular radius of 0.1\,$^{\circ}$ for each grid point.

The spectral \gray data from \citet{2006_HESS} is utilised, as this paper focussed on the TeV \gray sources in the inner part of the Galactic plane, including \hessj, providing a detailed look at the spectra and morphology of the TeV \gray observations. The \cite{2006_HESS} and \cite{2018_HGPS} data show good spectral matches as shown in \autoref{fig:2006_2018}. The \cite{2006_HESS} data provides more spectral data points which will allow the spectral shape of our model to be compared with observations. 
The \hessj spectral \gray observations from \citet{2006_HESS} were extracted from a circular region of ${\sim}0.36^{\circ}$ radius centred on \hessj ($l=8.4^{\circ}, b=-0.03^{\circ}$). 
We also make use of the spectral \gray data of \fhl from \citet{Fermi_3FHL_2017}, extracted from a disk region of $0.38^{\circ}$ (centred on $l=8.4^{\circ}, b=-0.09^{\circ}$).

\section{MODELLING}
\label{sec:mod_sim}
SNRs can be described as impulsive accelerators, in which CRs are accelerated by the SNR shock front, and escape into the ISM. We model the energy-dependent escape and subsequent diffusive transport of these particles using the solution to the isotropic diffusive transport equation \citep{1996_A&A}.  The injection of CRs is assumed to follow a power law, $\Ep^{-\alpha}$, with a spectral index of $\alpha$. Protons of different energies escape the accelerator at different times, where the higher energy protons leave the shock earlier than lower energy protons. Particles then diffuse through the ISM and interact to produce \grays. Where relevant, we assume either a Type Ia or Type II supernova explosion occurs with a total kinetic energy of $\ESN$, and an energy budget in CRs of $W_{\rm{p}}=\eta \ESN$, where $\eta \le 0.5$ \citep{Hadron_Ackermann_2013, Berezhko_Volk_1997, Volk_CRs_2000}.

\subsection{Proton flux}
The volume distribution of CRs, escaping from a shell, taking into account the time-dependent escape of protons of energy $\Ep$ is given by \cref{eqn:CR_distribution}.

\begin{equation}
\Jp(\Ep, R, t) = N_0  \Ep^{-\alpha} f_{\rm{p}}(\Ep, R, t) \ \ \rm{TeV^{-1}\,cm^{-3}},
\label{eqn:CR_distribution}
\end{equation}

where $f_{\rm{p}}(\Ep, R, t)$ is the probability density function (PDF) of protons, describing the probability of finding a particle of energy $\Ep$ at some distance from the accelerator, $R$. The time, $t$, is the time after the SN explosion. As we are interested in the evolutionary state of the SNR, we will use $t=t_{\rm{SNR}}$ in our model. The number density in our model is low ($n_{\rm{H}} < 10^2$\,cm$^{-3}$) which leads to a high proton-proton cooling time, $\tau_{\rm{pp}}\sim 6\times10^7 (n_{\rm{H}}/\rm{cm}^{-3})^{-1}$\,yr. As we consider relatively young accelerators ($\tau_{\rm{pp}} \gg t$, where $t<100$\,kyr), we can neglect the cooling term in our model.

We assume the energy budget in CRs is the total energy of particles with energies from $\Epmin$ to $\Epmax$, $W_p=N_0 \int_{\Epmin}^{\Epmax} E\,E^{-\alpha} dE$. From this definition we determine the normalisation factor, $N_0$, where the maximum energy is, $\Epmax$ and the minimum energy is $\Epmin=1$\,GeV. 

We note that Equation~3 from \citet{1996_A&A} relates to particles released from a point source. Our model instead, describes the time-dependent release of particles from a shell, so a modification (explained shortly) to account for this is adopted. The PDF of CR protons of energy $\Ep$ is given by:

\begin{equation}
    f_{\rm{p}}(\Ep, R, t) =
        \begin{cases}
            \fbubble(\Ep, R, t) & \Ep<\Epesc, R<\Resc \\
            \fdif(\Ep, R, t)  & \Ep>\Epesc, R>\Resc \\
            0 & \mathrm{Otherwise}
        \end{cases}
\label{eqn:full_PDF}
\end{equation}

The radius at which CR protons are released from the accelerator is given by \cref{eqn:r_esc} \citep{SNR_Reynolds_2008}. 
\begin{equation}
    \Resc = 0.31 
    \left( \dfrac{\ESN}{10^{51} \rm{erg}} \right) ^{1/5} 
    \left( \dfrac{n_0}{\rm{cm}^{-3}} \right) ^{-1/5} 
    \left(\dfrac{\tesc}{\rm{yr}}\right)^{2/5} \  \rm{pc} \, ,
    \label{eqn:r_esc}
\end{equation}

where $n_0$ is the ISM number density the SNR shock wave expands into \citep{CRs_Zirakash_2005, SNR_Reynolds_2008, n0_Ptuskin_2010}. The escape time of CR protons is \citep{Gabici_2009}:

\begin{equation}
    \tesc(\Ep) = \tsedov
    \left(
    \dfrac{\Ep}{\Epmax}
    \right) ^{-1/\deltap} \, ,
    \label{eqn:t_esc}
\end{equation}

where $\Epmax$ are the most energetic particles present at the start of the Sedov-Taylor phase. The onset of the Sedov-Taylor phase, $\tsedov$, is defined by \cref{eqn:t_sedov}.

The escape energy of protons is defined by rearranging \cref{eqn:t_esc} and setting $\tesc = \tsnr$:
 
\begin{equation}
\Epesc = \Epmax \left(\dfrac{\tsedov}{\tsnr} \right)^{\deltap} \, .
\label{eqn:E_sh}
\end{equation}

Particles at distance less than $\Resc$ and with energy less than $\Epesc$ are trapped inside a sphere, which we call the `bubble'. We assume that particles are uniformly distributed within this bubble, therefore the CR proton distribution can be described through:

\begin{equation}
\fbubble(\Ep, R, t) = \dfrac{1}{(4/3) \pi \Resc^3} \, .
\label{eqn:f_shell}
\end{equation}

The PDF for diffused CR protons is given by \citep{Mitchell_2021}:

\begin{equation}
\fdif(\Ep, R, t)  =
\dfrac{f_0}{\pi^{3/2}\Rdif^3}
\exp \left(- \dfrac{(R-\Resc)^2}{\Rdif^2} \right) \, .
\label{eqn:diff}
\end{equation}

We require our equation to be normalised, with a factor:

\begin{equation}
f_0 = \dfrac{\sqrt{\pi} \Rdif^3}{ (\sqrt{\pi}\Rdif^2 + 2\sqrt{\pi}\Resc^2)|\Rdif| + 4 \Resc \Rdif^2} \, .
\end{equation}

The diffusion radius is

\begin{equation}
\Rdif \equiv \Rdif(\Ep, t) = 2 \sqrt{D(\Ep)t'} \, ,
\label{eqn:R_dif}
\end{equation}

where $t' = t - \tesc(\Ep)$ represents the time the particles spend in the ISM. The energy-dependent diffusion coefficient from \citet{MCs_Gabici_2007} is used:

\begin{equation}
D(\Ep) = \chi D_0 \left( \dfrac{\Ep/\rm{GeV}}{B/3\mu \rm{G}}  \right)^{\delta} \, ,
\label{eqn:diff_coefficient}
\end{equation}

where $D_0$ takes the Galactic average value of $3\times10^{27}$\cms \citep{Berezinskii_1990}, the diffusion suppression factor is $\chi$ and the index of diffusion is $\delta$. 
The magnetic field is taken as a constant value of $B=B_0=10\,\mu$G as the average number density of the ISM surrounding \hessj is low \citep[$n_{\rm{H}}<300$\,cm$^{-3}$,][]{Mag_Crutcher_2010}. At low density, the magnetic field does not scale with density, due to diffuse clouds (low density) being assembled by motions along the magnetic field. The estimates from Zeeman splitting are also neglected at low density, as they are only significant for the dense ISM. This low magnetic field means particles in our model diffuse faster, as the diffusion coefficient increases.

\subsection{Gamma-ray flux}
The differential \gray flux (TeV$^{-1}$\,cm$^{-2}$\,s$^{-1}$) for the energy interval $(E_\gamma,\, E_\gamma+dE_\gamma)$ at position $R$ and time $t$ is computed through \citep{2006_Kelner}:

\begin{equation}
\mathcal{F_\gamma}(E_{\gamma}, R, t) =  \dfrac{N_{\rm{H}} A}{4\pi D^2}c\,
  \int\limits_{E_\gamma}^{\infty} \sigma_{\rm pp}(\Ep)\,
  \Jp(\Ep, R, t)\,
  F_\gamma\left(\dfrac{E_\gamma}{\Ep}, E_\gamma\right)\,
  \dfrac{d\Ep}{\Ep} \, ,
\label{eqn:diff_flux}
\end{equation}

where $A$ and $N_{\rm{H}}$ are the area, and total hydrogen column density of the region of interest, $D$ is the distance from Earth to the accelerator and $c$ is the speed of light. The inelastic cross section of proton-proton interactions is $\sigma_{\rm pp}$ (\cref{eqn:xsec_Kaf}) and $F_\gamma$ is the number of photons per collision given by \cref{eqn:gray_spec}.

\section{METHODOLOGY}
\label{sec:method}
\begin{figure}
\begin{center}
\includegraphics[width=0.88\columnwidth]{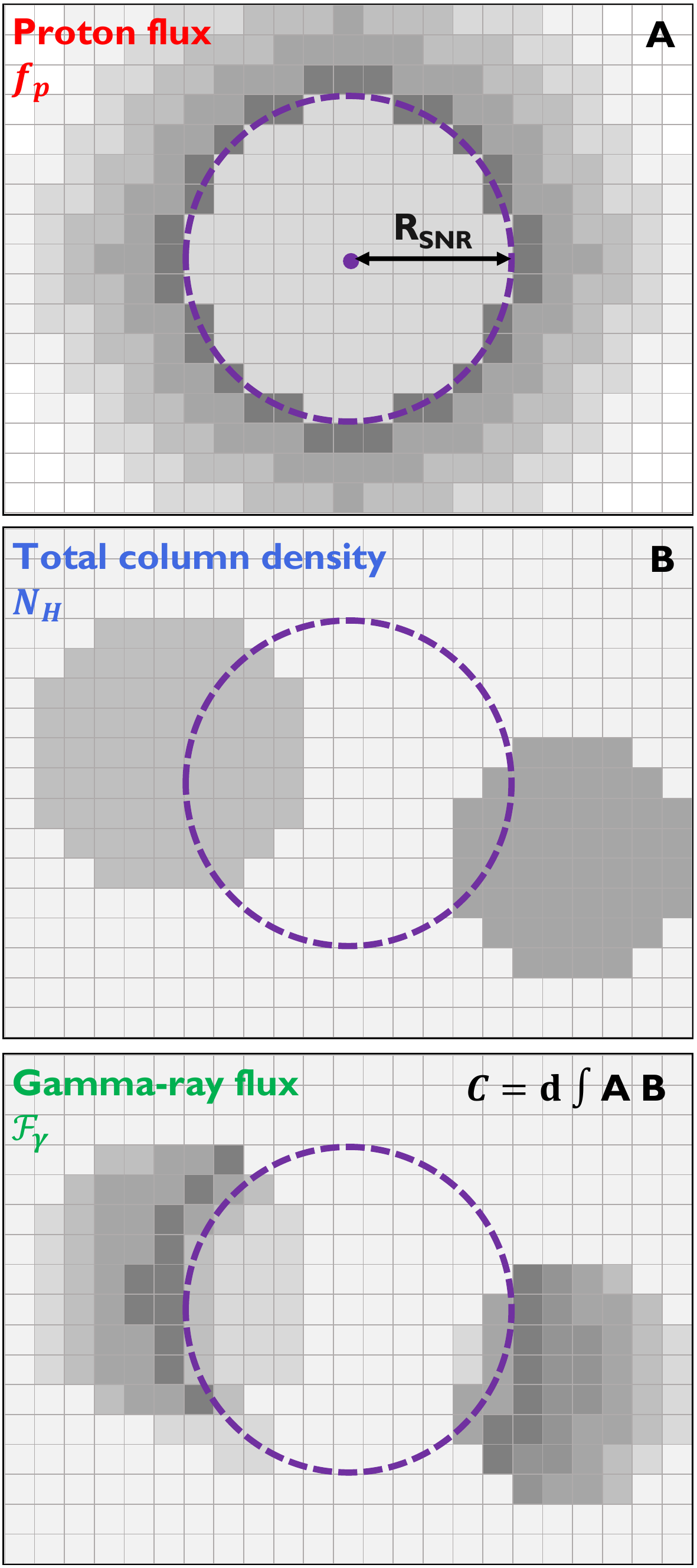}
\caption{Schematic illustrating our model. \textit{Top}: example distribution of CR protons (\cref{eqn:full_PDF}). \textit{Middle}: example distribution of the total column density gas. \textit{Bottom}: example distribution of \gray flux (\cref{eqn:diff_flux}) by integrating over the top and middle panels with some constant.}
\label{fig:model_schematic}
\end{center}
\end{figure}

We calculate the volume distribution of CR protons using \cref{eqn:CR_distribution} at every pixel in the total column density map, for a range of proton energies. The predicted 3D \gray map is created with the z-axis being \gray energy, via \cref{eqn:diff_flux}, by combining the proton map with the ISM distribution. The resulting \gray `cube' is used to extract spectra and integrated flux maps. \cref{fig:model_schematic} shows a schematic of the model.

Our model has a range of parameters. Multiple parameters have a similar effect on the model, for example, $\chi$ and $\delta$, which both effect the diffusion coefficient, which leads to a redundancy in our model solution. Due to this, we cannot perform a purely quantitative optimisation across the entire parameter space. Instead, we perform a systematic grid search over a range of model parameters, in which each combination is modelled, based on typical values from literature, as discussed below and compare these models to \gray observations. We calculate metrics to quantify the agreement of GeV-TeV observations with our model.
To compare the modelled morphology to the HGPS observations, the oversampling method from H.E.S.S. is applied, as described in \cref{sec:data}.

The spatial model of \grays is largely biased by the bubble component because the accelerators considered here lie within the extension of the \gray source. Therefore, we use the spectral model to determine the best matching model. We use the following metric for the spectral optimisation:
\begin{equation}
\dfrac{\chi^2}{m} = \left(\sum_i \dfrac{(O_i - F_i)^2}{G_i^2} \right) \dfrac{1}{m} \, ,
\label{eqn:chi_spectral}
\end{equation}

where $O_i$ is the observed flux, $F_i$ is the model flux, $G_i$ is the uncertainty in the observations, $i$ denotes each data point and $m$ is the number of data points. 
For testing the spatial agreement, we integrated the \gray cube from 1\,TeV to 100\,TeV to compare to the flux map from \citet{2018_HGPS}. As a metric for the spatial model we calculate the standard deviation, $S$ (\cref{eqn:std_spatial}), of the residuals but exclude the bubble region as we do not model the distribution of particles in detail there.

\begin{equation}
S = \dfrac{\sum_i (R_i - \mu )^2}{N}
\label{eqn:std_spatial}
\end{equation}

Here $R_i=O_i - F_i$ is the residual from the spatial morphology map (residual maps are provided in the supplementary material) for the $i^{\rm{th}}$ pixel, $\mu$ is the mean and $N$ is the number of pixels in the residual map.

\subsection*{Model parameter variation}
SNRs are thought to be the main source of CRs for energies below the knee of the CR spectrum \citep[at PeV energies,][]{Lucek_2000}. We define the maximum energy of protons, $\Epmax$, at the start of the Sedov phase to be either 1\,PeV or 5\,PeV \citep{Gabici_2009}, for \cref{eqn:t_esc,eqn:E_sh} and in the normalisation factor, $N_0$. The energy budget, $\Wp$, is a free parameter which is optimised by minimising the residual between the observations and model spectra. From diffusive shock acceleration theory, we expect a power law with a spectral index of $\alpha \approx 2$ \citep{Malkov_Drury_2001}. Therefore, we vary the spectral index from 1.8 to 2.4 in our model. 

The factor $\chi$ varies from $0.001$ to $1$, the lower limit ($\chi=0.001$) is potentially applicable to the dense regions of interstellar gas that the CRs may diffuse through, and the upper limit is taken from various observations \citep[see][and references therein]{Feijen_2020}. Typically, $\delta$ varies from $0.3$ to $0.7$ \citep{Berezinskii_1990} to allow for a range of turbulent spectra to be investigated. Here $0.3$ corresponds to Kolmogorov turbulence (indicating slower diffusion), $0.5$ indicates Kraichnan turbulence \citep{Strong_2007} and $0.7$ is consistent with a good fit to the Boron to Carbon ratio measurements.

In \cref{eqn:t_esc}, $\deltap$ describes the energy-dependent release of CRs.  The lower limit is taken to be $1/5$ which is for a simple stationary particle and the upper limit is taken to be 2.5 \citep{CRs_Zirakash_2005, SNRs_Celli_2019, Gabici_2009}. For the Sedov time in \cref{eqn:t_sedov} we take the typical mass and energy values for different supernovae types. We find for Type Ia, where $\Mej=1M_\odot$ \citep{CRs_Zirakash_2005} and $\ESN=10^{51}$\,erg, the Sedov time is $\tsedov \sim 230$yr. For Type II, where $\Mej=10M_\odot$  and $\Mej=20M_\odot$ \citep{Mej_Heger_2003} with $\ESN=10^{51}$\,erg, the Sedov time is $\tsedov\sim 1600$yr and $\tsedov\sim 2850$yr, respectively. Type II can also have a higher total kinetic energy \citep{2004_energy_SN} of $\ESN=10^{52}$\,erg for $\Mej=10M_\odot$ and $\Mej=20M_\odot$, where the Sedov time is $\tsedov\sim 500$yr and $\tsedov\sim 900$yr, respectively.

We expect $n_0$, from \cref{eqn:r_esc}, to be a small value as it is close to `SNR birth' before the shock wave has interacted with the gas. If we know the age, $\tsnr$, and radius, $\rsnr$, of the SNR we estimate $n_0$ by rearranging \cref{eqn:r_esc}:
\begin{equation}
n_0 = \left[ \dfrac{0.31 \rm{pc}}{\rsnr} \left( \dfrac{\ESN}{10^{51}\rm{erg}} \right) ^{1/5}
\left(\dfrac{\tsnr}{\rm{yr}}\right)^{2/5}   \right]^5 \  \rm{cm^{-3}} \, .
\end{equation}

If $\rsnr$ is not known, $n_0$ takes on values from 0.1 to 20\,cm$^{-3}$. The model parameters discussed are summarised in \autoref{tab:model_params}.

\begin{table}
\begin{center}
\caption{Model parameters with their discrete values. The spectral index is $\alpha$, the diffusion suppression factor and the index of diffusion coefficient are given by $\chi$ and $\delta$, respectively. 
The maximum CR proton energy is $\Epmax$ and $\deltap$ describes the energy-dependent release of CRs.
$\ESN$  is the kinetic energy released at the supernova explosion and $\Mej$ is the mass of the ejecta.}
\begin{tabular}{cc}
\hline 
Parameter      &   Values \\ 
\hline
  $\alpha$           &  1.8, 2.0, 2.2, 2.4 \\
  $\chi$             &  0.001, 0.01, 0.1, 1.0 \\
  $\delta$           &  0.3, 0.4, 0.5, 0.6, 0.7 \\
  $\deltap$         &  0.2, 1.4, 2.5 \\
$\Epmax$ &  1, 5\,PeV \\
\hline
                     &  1$M_\odot$ and $10^{51}$\,erg (Type Ia) \\
$\Mej$ and $\ESN$  &  10$M_\odot$, 20$M_\odot$ and $10^{51}$\,erg (Type II) \\
                     &  10$M_\odot$, 20$M_\odot$ and $10^{52}$\,erg (Type II)  \\
  $n_0$*             &  0.1, 1, 10, 20\,cm$^{-3}$ \\
\hline
\end{tabular}
\\ \small * For progenitor SNR of \psrj
\label{tab:model_params}
\end{center}
\end{table}

\section{BEST MATCHING MODELS}
\label{sec:results}
The following section considers the best matching models for each accelerator, based on the parameter space and minimising the spectral and spatial criteria (\cref{eqn:chi_spectral,eqn:std_spatial}, respectively), as described in \cref{sec:method}. Specifically, the 5 best matching spectral models are chosen for each accelerator.

\subsection{\snrg}
\snrg is believed to be contained in Component 2 (\vlsr=\ 30\ to\ 40\,\kms). We test the various model parameters for both Type Ia and Type II supernovae for both suggested ages of \snrg, 15\,kyr and 28\,kyr, the results shown here are the closest matching spectra to the observations. \Cref{tab:G8_15kyr,tab:G8_28kyr} show the 5 best matching spectral models with their ranking parameters for the spectral and spatial models, $\chi^2/m$ and $S$ respectively.

\begin{figure}
\includegraphics[width=\columnwidth]{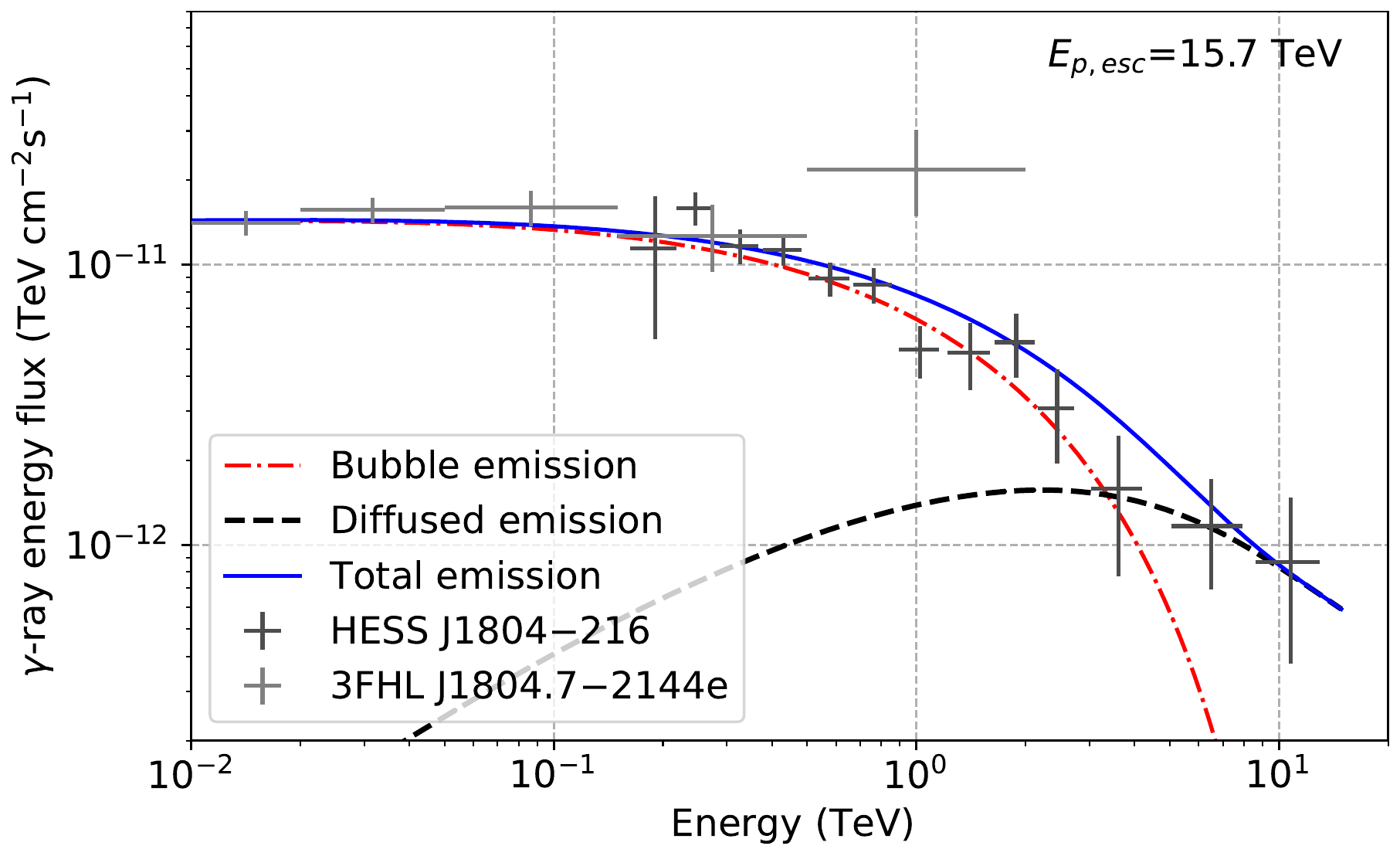}
\includegraphics[width=\columnwidth]{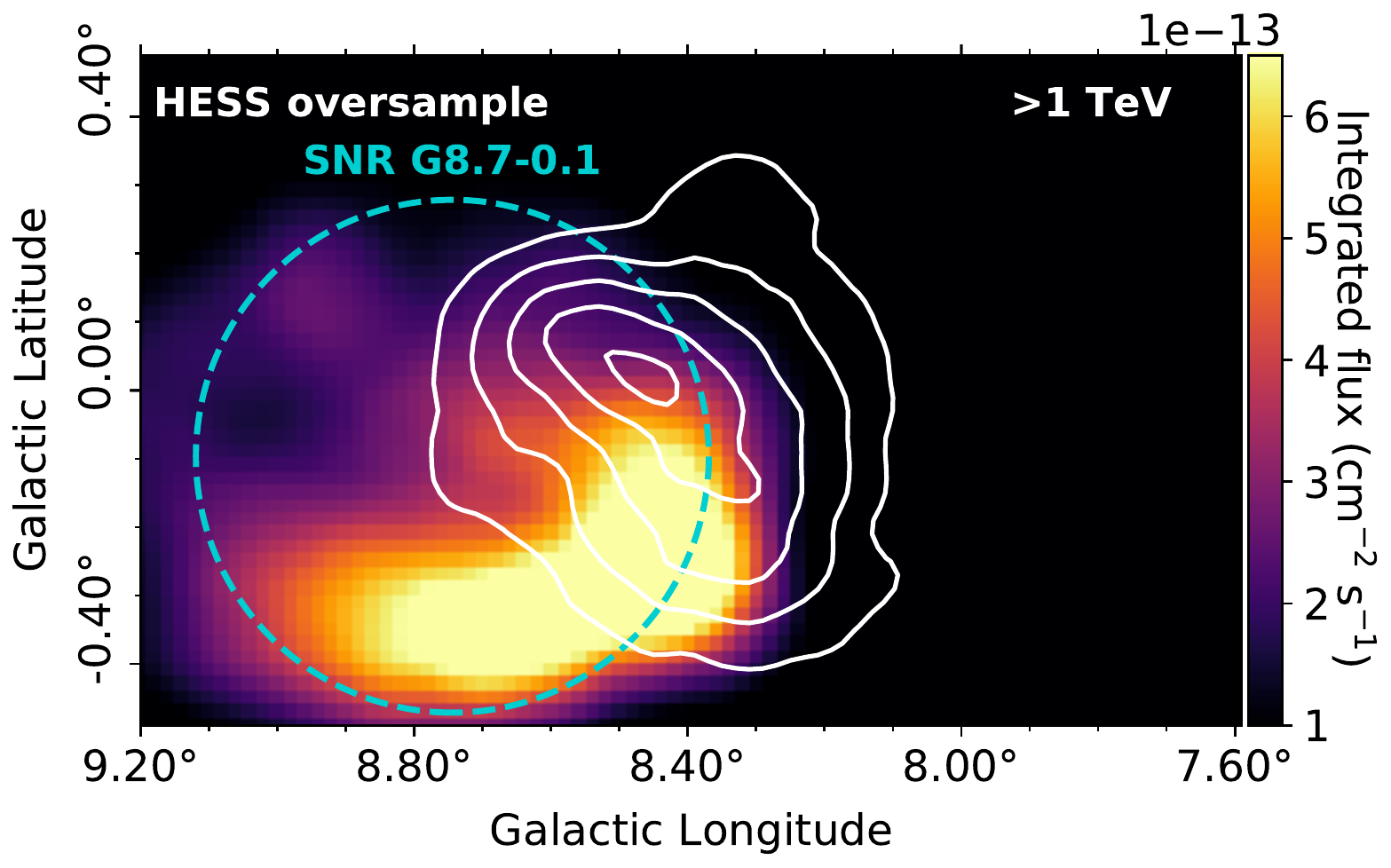}
\caption{Best matching spectral and spatial model (G4a) for \snrg with an age of 15\,kyr. \textit{Top}: \gray spectral model shown in blue. The diffused and bubble spectra are shown by the dashed black curve and dot-dashed red curve, respectively, with \hessj observations in dark gray and \fhl observations in light gray. \textit{Bottom}: \gray flux map above 1\,TeV towards \hessj. The escape radius, $\Resc$, is shown by cyan dashed circle. The TeV \gray emission from \citet{2018_HGPS} shown by the solid white contours ($2\times10^{-13}$ to $6\times10^{-13}$\,cm$^{-2}$\,s$^{-1}$ levels). The lower limit of the colorbar is set to $10^{-13}$\,cm$^{-2}$\,s$^{-1}$ to exclude any emission below the sensitivity of H.E.S.S..
Model parameters: $\alpha=2.0$, $\chi=0.01$, $\delta=0.6$, $\deltap=2.5$, $\Epmax=1$\,PeV, Type II SN, $\Epesc=15.8$\,TeV. For the spectral model $\chi^2/m=1.1$.}
\label{fig:G8_15_best}
\end{figure}

\begin{figure}
\includegraphics[width=\columnwidth]{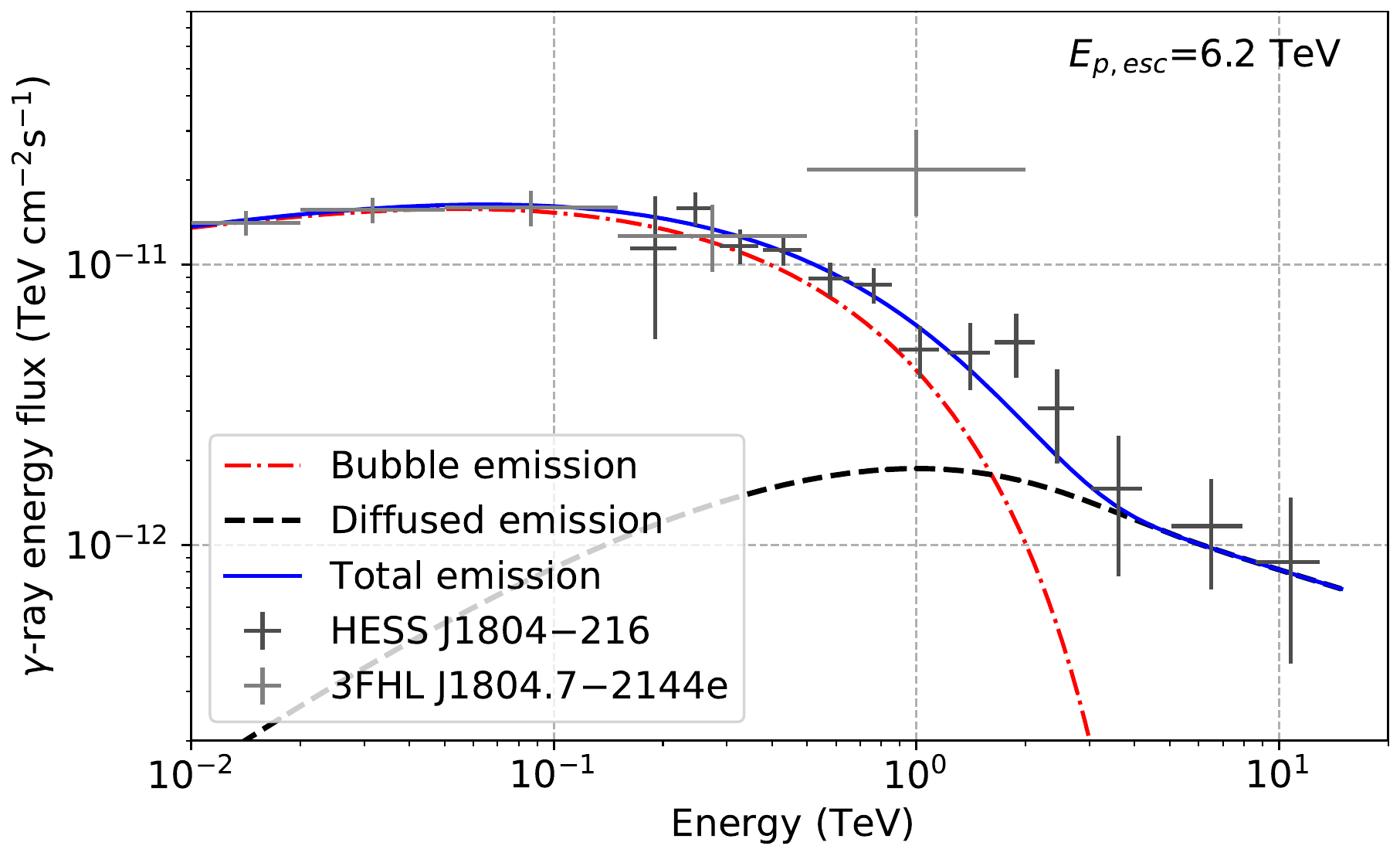}
\includegraphics[width=\columnwidth]{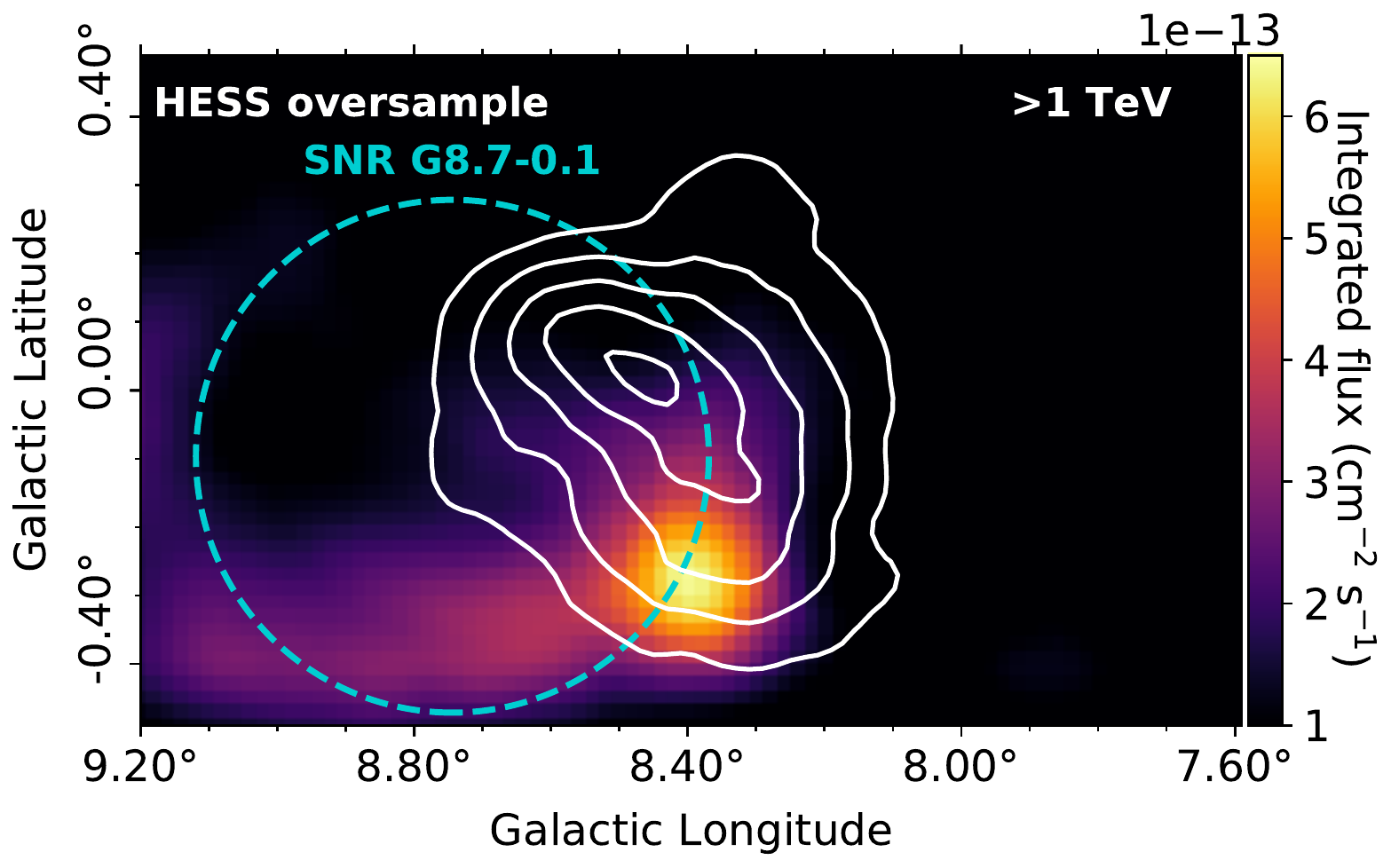}
\caption{Best matching spectral and spatial model (G2b) for \snrg with an age of 28\,kyr.  \textit{Top}: \gray spectral model shown in blue. The diffused and bubble spectra are shown by the dashed black curve and dot-dashed red curve, respectively, with \hessj observations in dark gray and \fhl observations in light gray. \textit{Bottom}: \gray flux map above 1\,TeV towards \hessj. The escape radius, $\Resc$, is shown by cyan dashed circle. The TeV \gray emission from \citet{2018_HGPS} shown by the solid white contours ($2\times10^{-13}$ to $6\times10^{-13}$\,cm$^{-2}$\,s$^{-1}$ levels).
Model parameters: $\alpha=1.8$, $\chi=0.1$, $\delta=0.4$, $\deltap=1.4$, $\Epmax=5$\,PeV, Type Ia SN, $\Epesc=6.2$\,TeV. For the spectral model $\chi^2/m=0.8$.}
\label{fig:G8_28_best}
\end{figure}

The 5 best matching models for \snrg for each age show a range of $\chi$ and $\delta$ values. Typically a moderately slow diffusion with $\chi=0.001$\,or\,$0.01$ is seen, which is consistent with other studies in which the diffusion coefficient is suppressed \citep{MCs_Gabici_2007, Li_Chen_2010, Chi_Giuliani_2010}. The index for the energy-dependent release of CRs, $\deltap$, takes on values of 1.4 or 2.5. A mixture of the $\Epmax$ values are present. 
For \snrg we find both a Type Ia SN and Type II SN, match the \gray spectra well for both ages of this accelerator. The escape energy of protons, $\Epesc$, in \cref{fig:G8_15_best,fig:G8_28_best} are $15.8$\,TeV and $6.2$\,TeV respectively. CR protons with energy lower than $\Epesc$ are still confined in the bubble, however higher energy particles have escaped. This is shown by splitting the spectra into its bubble and diffused components as depicted in the top panels of \cref{fig:G8_15_best,fig:G8_28_best}.

For our best matching models, \cref{fig:G8_15_best,fig:G8_28_best}, the spectra tend to match well at low energies. The escape energies for these spectra are quite high, therefore the model spectra are largely dominated by the bubble component. This is seen at higher energies, where the model begins to deviate from the observations, particularly in \autoref{fig:G8_15_best}.

The spatial morphology above 1\,TeV cannot explain the \gray emission from \hessj. One contributing factor is the bubble component encompassing a large area of the H.E.S.S. source. The simulated \gray emission shows a peak toward the northern edge of \hessj and a lack of \gray emission towards the western TeV peak of \hessj. Simulated \gray emission is also present at the outer western edge of \snrg for both ages, which is not present in the observations from H.E.S.S.. The closest TeV \gray source from the HGPS is HESS\,J1808$-$204, which is an extended source located at $l=10.01$, $b=-0.24$, which is not close enough to \snrg to provide the \gray emission at this position.

\subsection{Progenitor SNR of \psrj}
The progenitor SNR is assumed to have an age of 16\,kyr, as per the age of \psrj, and is placed at the birth position of \psrj as shown in \autoref{fig:HESS_flux}. The progenitor SNR is believed to be in Component 1 (\vlsr=\ 20\ to\ 30\,\kms). We test a range of combinations of model parameters for a core-collapse supernova, as a pulsar is attached to the system. The 5 best matching spectral models with their ranking parameters for the spectral and spatial models, $\chi^2/m$ and $S$ respectively, are shown in \autoref{tab:PJ_16kyr}.

\begin{figure}
\includegraphics[width=\columnwidth]{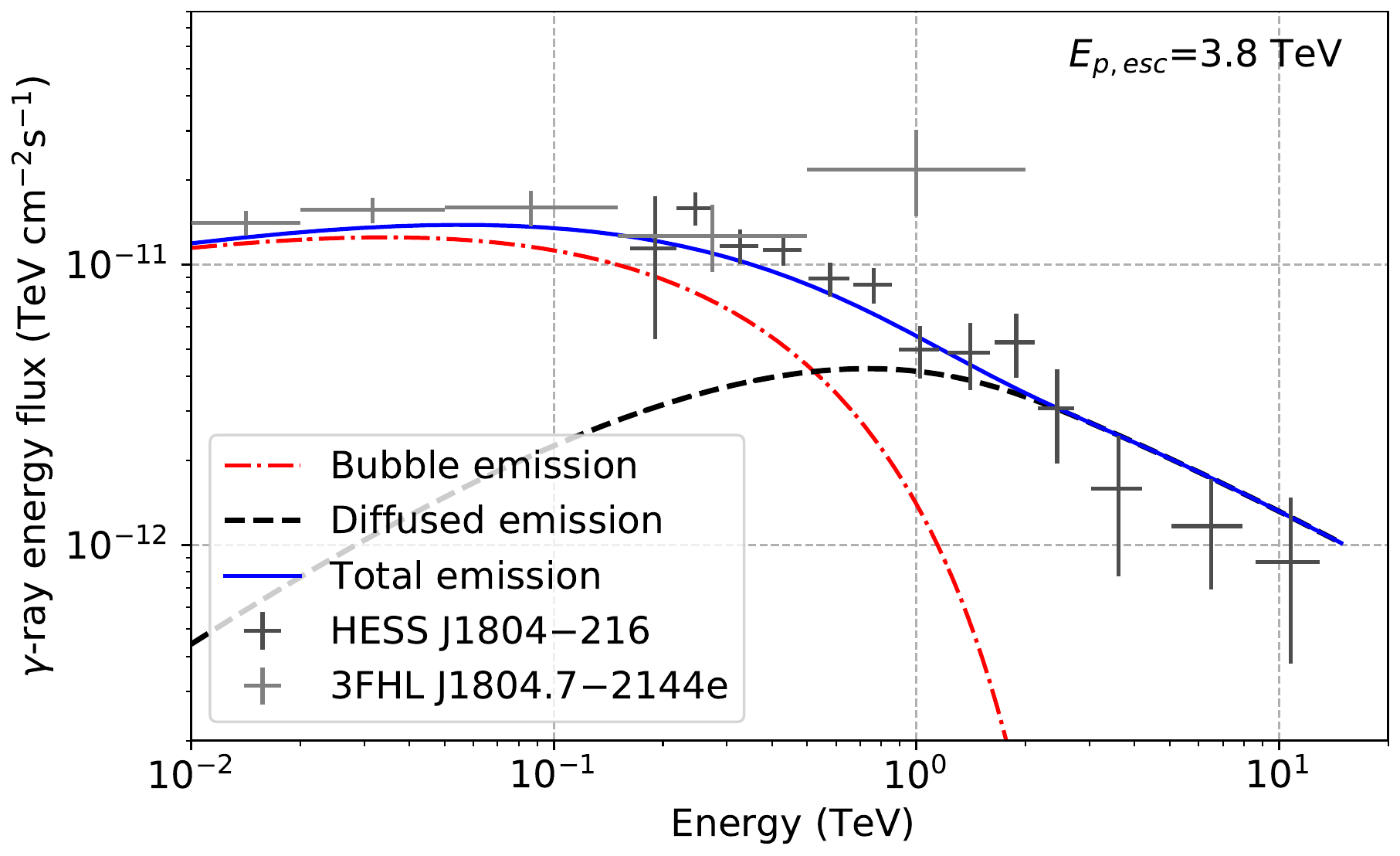}
\includegraphics[width=\columnwidth]{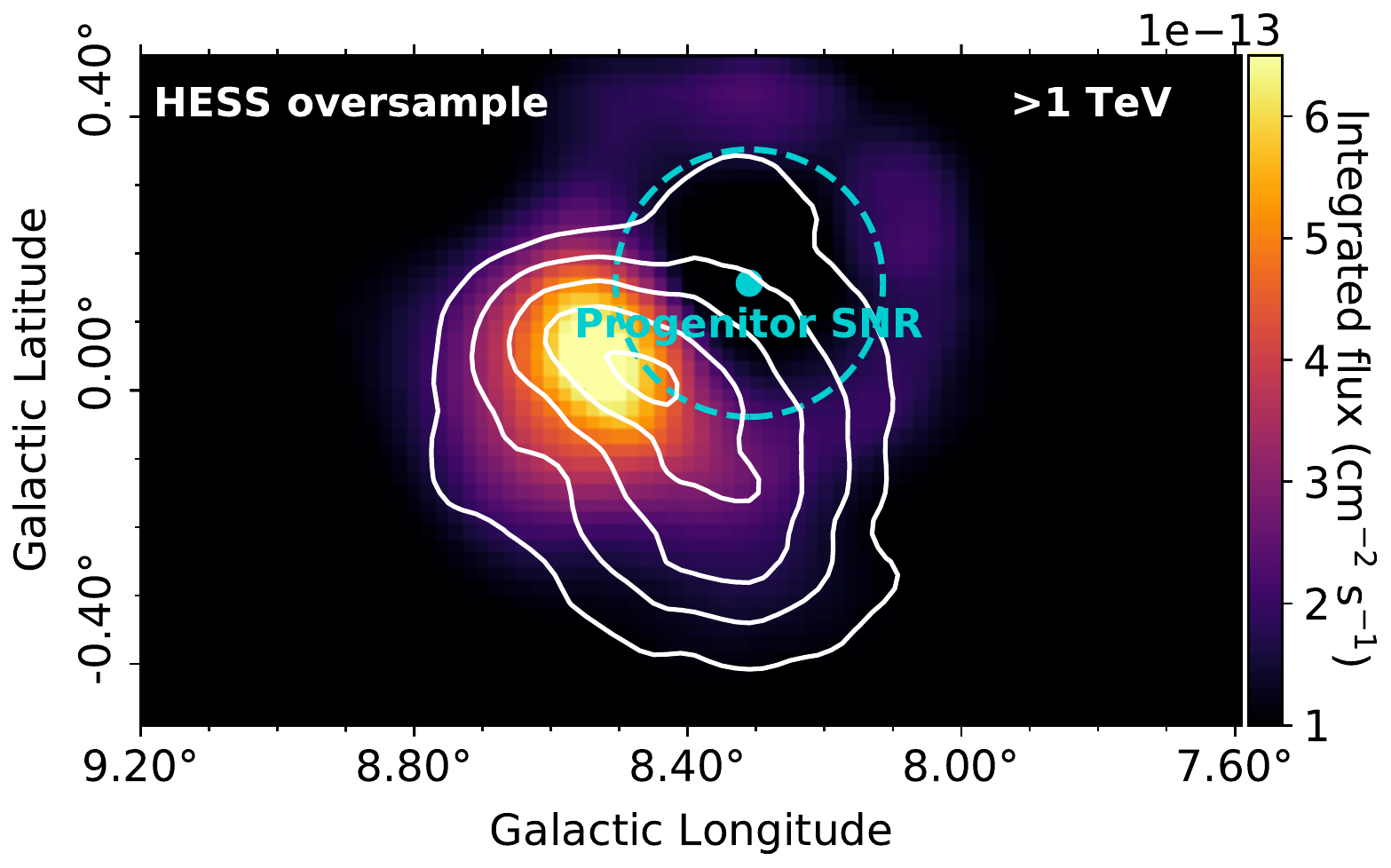}
\caption{Best matching spectral and spatial model (P1) for the progenitor SNR of \psrj. \textit{Top}: \gray spectral model shown in blue. The diffused and bubble spectra are shown by the dashed black curve and dot-dashed red curve, respectively, with \hessj observations in dark gray and \fhl observations in light gray. \textit{Bottom}: \gray flux map above 1\,TeV towards \hessj. The progenitor SNR of \psrj is indicated by the cyan dot and the escape radius, $\Resc$, is shown by cyan dashed circle. The TeV \gray emission from \citet{2018_HGPS} shown by the solid white contours ($2\times10^{-13}$ to $6\times10^{-13}$\,cm$^{-2}$\,s$^{-1}$ levels). Model parameters: $\alpha=1.8$, $\chi=0.1$, $\delta=0.4$, $\deltap=1.4$, $\Epmax=5$\,PeV, Type Ia SN, $\Epesc=6.2$\,TeV. For the spectral model $\chi^2/m=0.8$.}
\label{fig:PJ_best}
\end{figure} 

\autoref{tab:PJ_16kyr} shows the 5 best matching spectral models typically have a spectral index of $\alpha=1.8$, with $\chi=0.001$ or $0.01$ and a range of $\delta$. Similarly to \snrg, this indicates moderately slow diffusion of particles. Both values of the maximum energy, $\Epmax$, are present and $n_0$ takes on all values in the parameter space. The total kinetic energy, $\ESN$, is typically the higher value from our parameter space of $10^{52}$\,erg, with the ejecta mass being either $10 M_\odot$ or $20 M_\odot$. The index for the energy-dependent release of CRs $\deltap=2.5$, the highest value chosen in our parameter space. The escape energy for the top model is $\Epesc=3.8$\,TeV, therefore some particles are still trapped in the bubble and some have diffused, this can be seen in the spectral components in \autoref{fig:PJ_best}.

For this accelerator, both the spectral and spatial morphology match the observations well. The modelled integrated \gray map peaks towards the northern TeV peak of \hessj from \citet{2018_HGPS}. There is weaker modelled \gray emission overlapping the entire \gray source and a lack of modelled emission outside the \hessj region. No strong \gray emission is present outside \hessj unlike with \snrg. However, parts of the morphology do not match well due to the bubble component and the lack of \gray emission in the southern and eastern edges of \hessj.

\section{DISCUSSION}
\label{sec:discussion}
In the previous section we compared the model emission with observational \gray emission above 1\,TeV. Fig.\,17 from \citet{2006_HESS} shows the morphology of \hessj above 0.2\,TeV, which is similar to the integrated flux morphology above 1\,TeV in \autoref{fig:HESS_flux} \citep{2018_HGPS}. 
Extensions of both \hessj and \fhl overlap (\autoref{fig:HESS_flux}). To probe the effects of the bubble and diffused components we look at the \gray emission in different energy bands: $E_{\gamma}{=}10{-}100$\,GeV, $E_{\gamma}{=}0.1{-}1$\,TeV, $E_{\gamma}{=}1{-}10$\,TeV, and $E_{\gamma}{=}10{-}100$\,TeV. 

\begin{figure}
\includegraphics[width=\columnwidth]{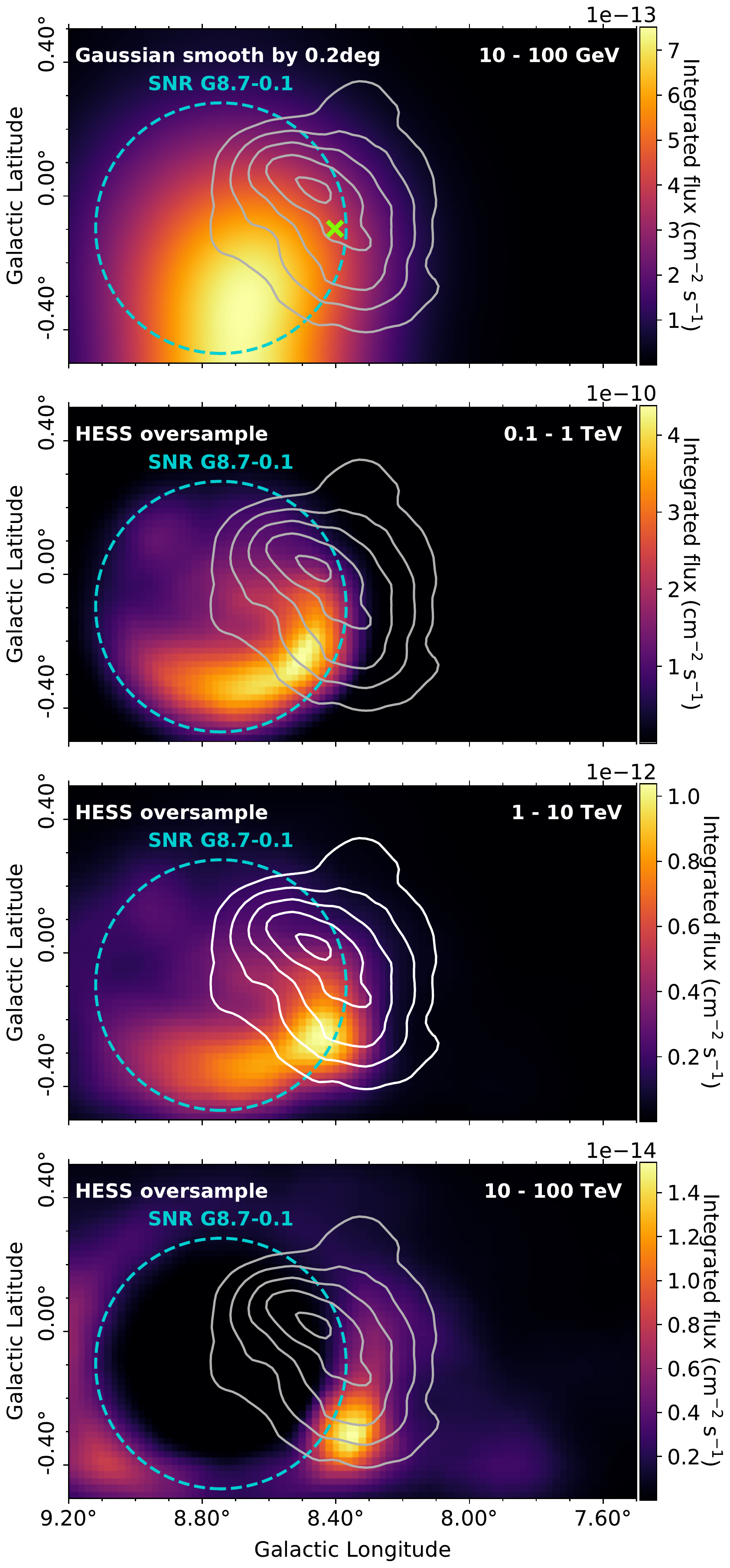}
\caption{\gray flux maps for various energy bands towards \hessj for the best matching model (G4a) for \snrg with an age of 15\,kyr. The escape radius, $\Resc$, is shown by cyan dashed circle. The green cross indicates the centroid of \fhl. The first energy band ($E_{\gamma}{=}10{-}100$\,GeV) corresponds to energies detected by \textit{Fermi}-LAT so is smoothed using the \textit{Fermi}-LAT PSF of ${\sim} 0.2^{\circ}$ above 10\,GeV \citep{2017_Pass8}. The other energy bands are oversampled using the H.E.S.S. method described in \cref{sec:data}. The TeV \gray emission above 1\,TeV from \citet{2018_HGPS} is shown by the solid white contours in the third panel ($2\times10^{-13}$ to $6\times10^{-13}$\,cm$^{-2}$\,s$^{-1}$ levels) and by grey contours in the other panels as a reference for where the \gray emission is expected. 
Model parameters as described in \autoref{fig:G8_15_best}.}
\label{fig:G8_15_spatial_best}
\end{figure} 

\begin{figure}
\includegraphics[width=\columnwidth]{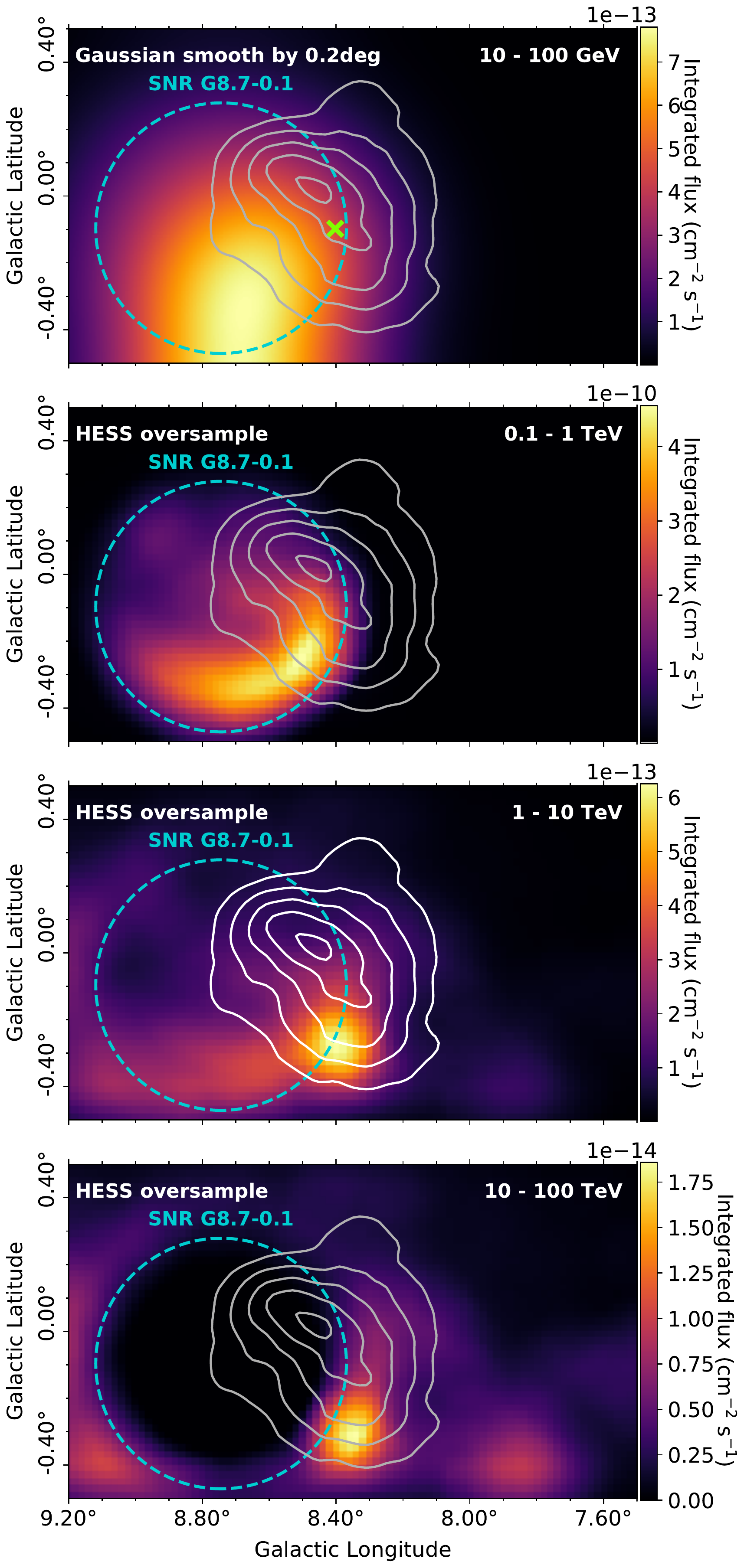}
\caption{\gray flux maps for various energy bands towards \hessj for the best matching model (G2b) for \snrg with an age of 28\,kyr. The escape radius, $\Resc$, is shown by cyan dashed circle. The green cross indicates the centroid of \fhl. The solid white/grey contours are as described in \autoref{fig:G8_15_spatial_best}. Model parameters as described in \autoref{fig:G8_28_best}.}
\label{fig:G8_28_spatial_best}
\end{figure} 

The morphology of the different energy bands for the \snrg accelerator show that the bubble component provides stronger \gray emission compared to the diffused component (as shown in \cref{fig:G8_15_spatial_best,fig:G8_28_spatial_best}). At higher energies ($E_{\gamma}{=}1{-}10$\,TeV, and $E_{\gamma}{=}10{-}100$\,TeV) the model exhibits strong emission towards the southern edge of \hessj, which does not overlap with the TeV peak.
In comparing the four energy bands to morphology of \hessj, it is clear the morphology is quite different.

\begin{figure}
\includegraphics[width=\columnwidth]{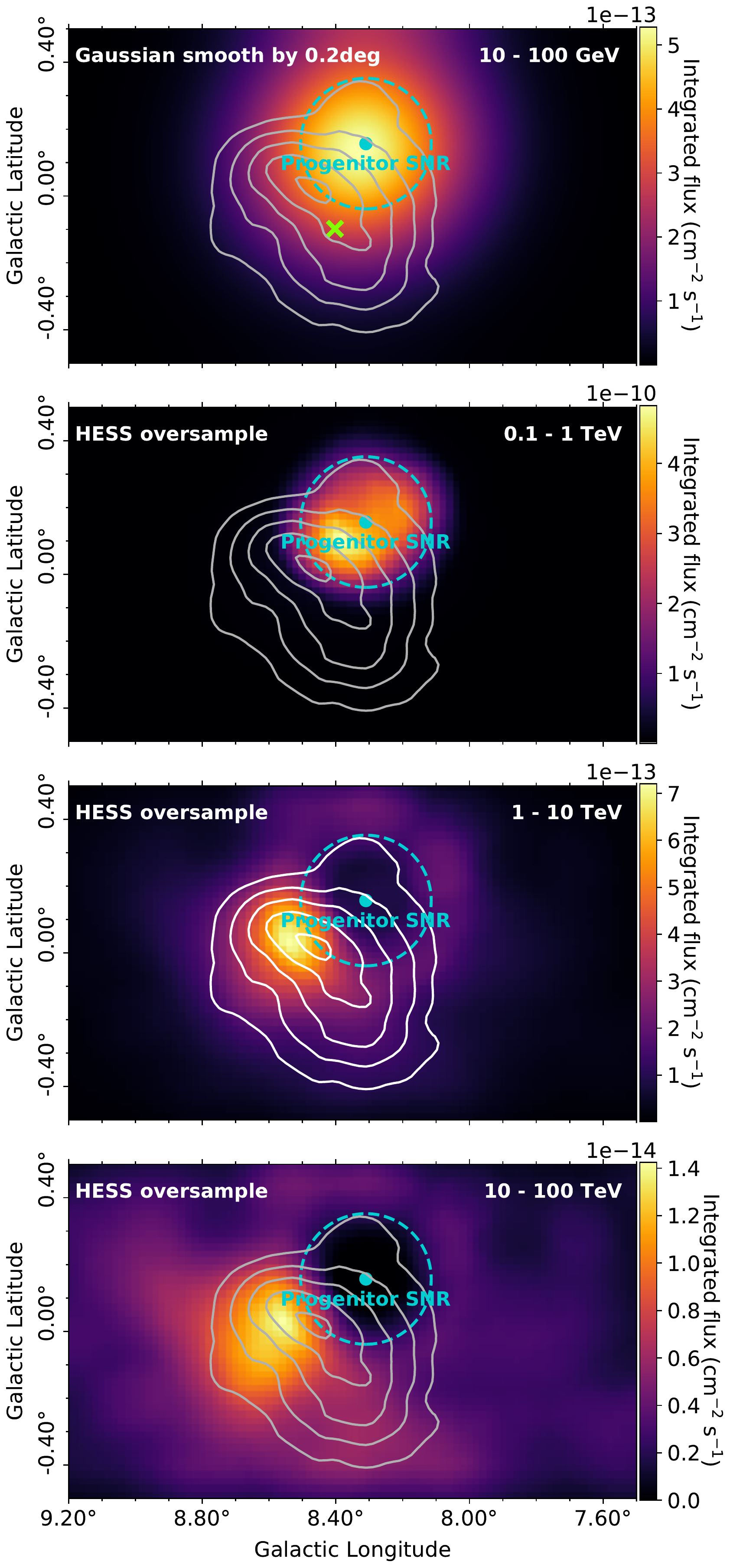}
\caption{\gray flux maps for various energy bands towards \hessj for the best matching model (P1) for the progenitor SNR of \psrj. The escape radius, $\Resc$, is shown by cyan dashed circle. The green cross indicates the centroid of \fhl. The solid white/grey contours are as described in \autoref{fig:G8_15_spatial_best}. Model parameters as described in \autoref{fig:PJ_best}.}
\label{fig:PJ_16_spatial_best}
\end{figure}

\autoref{fig:PJ_16_spatial_best} shows the morphology from different energy bands for the best model of the progenitor SNR. The bubble tends to dominate at the two lower energy bands ($E_{\gamma}{=}10{-}100$\,GeV and $E_{\gamma}{=}0.1{-}1$\,TeV), with strong emission towards the pulsar birth position. The two higher energy bands ($E_{\gamma}{=}1{-}10$\,TeV and $E_{\gamma}{=}10{-}100$\,TeV) show the \gray peak moves away from the pulsar position and closer to the TeV \gray peak from HGPS. The emission in these bands becomes diffusion-dominated with extended emission outside the bubble. 

Overall, the \gray emission in the lower energy bands of \cref{fig:G8_15_spatial_best,fig:G8_28_spatial_best,fig:PJ_16_spatial_best} are dominated by the bubble component as the particles have energy less than the escape energy, so are still confined to the bubble region. However, above 1\,TeV the diffused component becomes dominant.
The morphology of the lower energy band can be compared with the spatial morphology of \textit{Fermi}-LAT observations. However, \textit{Fermi}-LAT has poor resolution (with a PSF of 0.2$^{\circ}$ above 10\,GeV), so the spatial map from \textit{Fermi}-LAT observations is described only by a circular region. 
Therefore, we can only compare the model morphology with the position and extent of \fhl (seen in \autoref{fig:HESS_flux}). For each accelerator model the peak of \gray emission is offset from the 3FHL catalogue position (by up to 0.4$^{\circ}$).

There are a number of known limitations in our model. These arise from:
\begin{enumerate}
  \item 2D propagation of CR protons
  \item 2D arrangement of the ISM
  \item Uniform distribution of CRs inside bubble
  \item Simple assumption of ISM inside bubble
  \item Spherical uniformity of the SNR evolution
\end{enumerate}

The main limitation comes from assuming a 2D geometry for the proton model in addition to the accelerators lying within the \gray source. This 2D approach to modelling the \grays has previously been implemented e.g. \citet{Casanova_2010}, in which the accelerator is further away from the \gray source. 
In our model, we have a special case in which both accelerators lie within the extension of the \gray source. As we do not model the diffusion inside the bubble component, our results are biased.
In our 2D proton model, we do not consider particles diffusing along our line of sight, therefore we are not taking into account the emission foreground/background to the SNR bubble. 
This makes reconciling the morphology inside the bubble with the observations difficult and is the reason we do not use the spatial criterion (\cref{eqn:std_spatial}) to test which model performs best.

Another limitation comes from the assumptions regarding the ISM surrounding \hessj. We assume a 2D model in which the brightness temperature gas cube is integrated over to obtain a total column density map. By using the column density map we assume that all gas we integrated over is interacting with the CRs. Due to this, we could be including gas that is foreground and/or background to the accelerator, which may not be physically able to interact with the accelerated protons. 
This effect is more prominent for the bubble component which has radius $\Resc{\sim}30$\,pc compared to the entire column which is integrated over 10\kms corresponding to ${\sim}800$\,pc from the GRC. Our model could therefore be over predicting the bubble component. 

One of the largest uncertainties in our model comes from the distribution of CRs and density of the ISM inside the bubble. We assume the CRs are uniformly distributed in the bubble \citep[see e.g. ][]{Zirakash_uniform_2010}, however realistically the particle distribution is more complex, with CRs likely accumulating in the shock region (i.e. the expanding SNR shell), as explored by some theoretical studies \citep{CRs_Zirakash_2005, SNRs_Celli_2019, 2020_Brose_SNRs}. They suggest the CR distribution can be `shell’ brightened. 
From diffusive shock acceleration theory, it is possible that the shocked ISM could be $\sim$4 times denser than the unshocked gas \citep{SNR_Reynolds_2008}, according to the expected shock compression factor. Furthermore, it is possible for some of the ISM here to be dissociated by the SNR shock, thereby reducing the density \citep{Fukui_density_2003,Sano_density_2020}. 

For limitation (v) we assume a constant number density for the SNR evolution ($n_0$, in \cref{eqn:r_esc}), leading to spherical uniformity of the particle escape radius. The escape energy is dependent upon $\deltap$, which describes the energy dependent release of CRs, which can also effect the evolution of the SNR. A future version of the model would involve tracing the evolution of the shock in closer detail, including the escape radius as it changed with ISM density.

\section{CONCLUSIONS}
\label{sec:conclusion}
We developed a model to investigate the distribution of \grays towards \hessj for two SNRs in the hadronic scenario. This is a first attempt to model the morphology of \grays towards \hessj.
\gray spectra and morphology maps of \snrg and the progenitor SNR of \psrj were generated for a range of model parameters and compared to observations to gain an understanding of the origin of \hessj. It was found that the progenitor SNR is the most promising candidate to be creating the TeV \grays, however, we are limited by the bubble component. The modelled \gray morphology from \snrg does not match the \gray morphology from observations well, therefore it is either only a minor contributor or does not contribute to the observed \gray emission.

\begin{figure}
\includegraphics[width=\columnwidth]{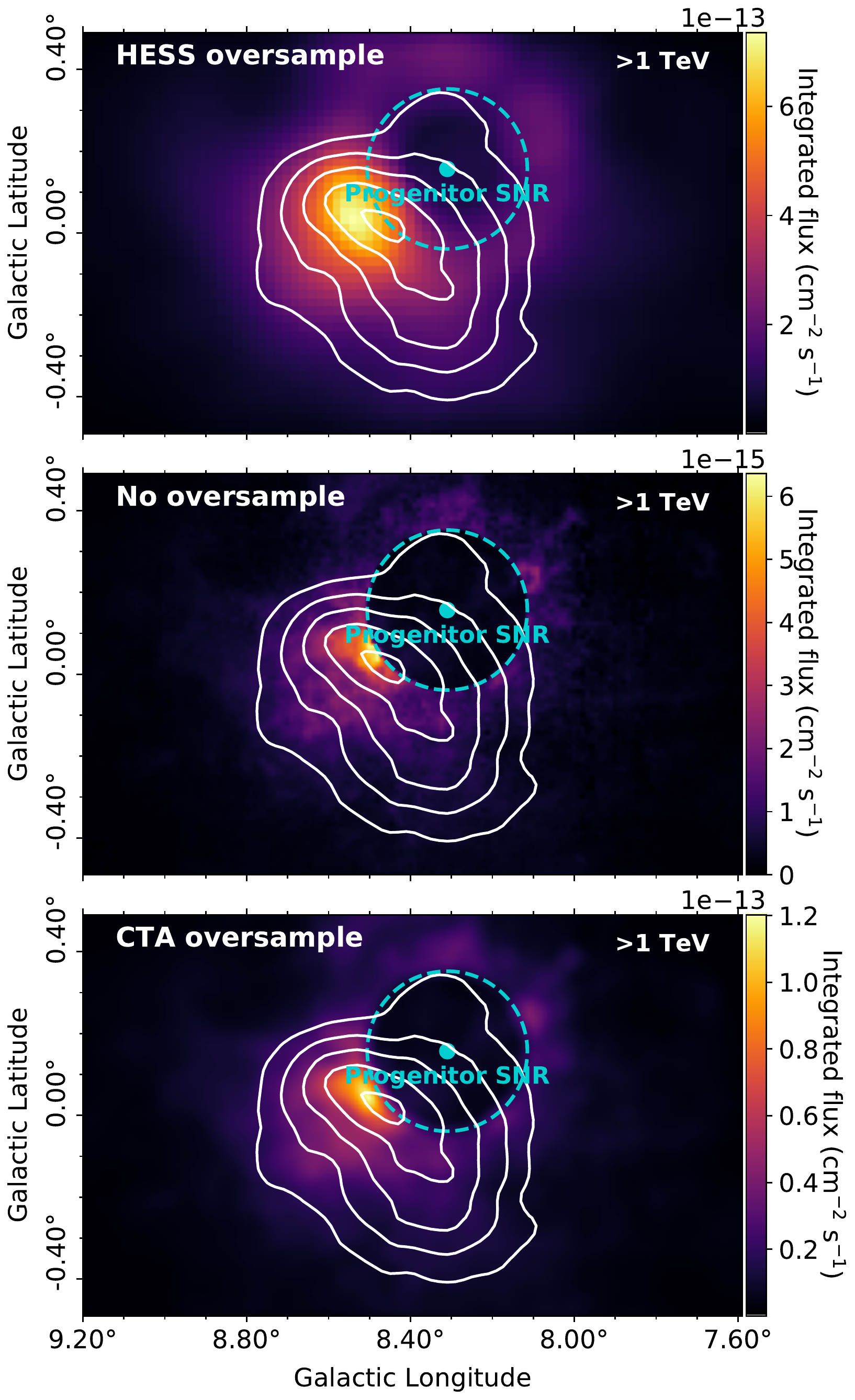}
\caption{\gray flux maps above 1\,TeV towards \hessj with the progenitor SNR of \psrj as the accelerator (for the top model, P1). The TeV \gray emission from \citet{2018_HGPS} shown by the solid white contours ($2\times10^{-13}$ to $6\times10^{-13}$\,cm$^{-2}$\,s$^{-1}$ levels). \textit{Top}: Map oversampled with H.E.S.S. factor of radius=$0.1^{\circ}$ with grid size of $0.02^{\circ}$. \textit{Middle}: Map not oversampled, with same resolution as gas map. \textit{Bottom}: Map oversampled with CTA factor of radius=$0.03^{\circ}$ with grid size of $0.01^{\circ}$.}
\label{fig:CTA}
\end{figure}

The \gray observatory CTA (Cherenkov Telescope Array) aims to improve the current measurements from other Imaging Atmospheric Cherenkov Telescope. More detailed features in the morphology may be resolved with CTA, which will provide unprecedented angular resolution and sensitivity. The angular resolution of CTA will reach a few arcmins, comparable to the angular resolution of the Mopra radio telescope, which is utilised for our gas measurements. Here we compare the best matching model of the progenitor SNR of \psrj (P1) for three different angular resolutions using different oversampling settings for each instrument. The original model (no oversampling, same resolution as the gas maps), the H.E.S.S. oversample (radius=$0.1^{\circ}$ with a grid size of $0.02^{\circ}$) and the expected CTA oversample \citep[radius=$0.03^{\circ}$ with a grid size of $0.01^{\circ}$, as per the expected angular resolution from][]{CTA_Science_2019} maps are shown in \autoref{fig:CTA}. 

\autoref{fig:CTA} shows that the oversampling of CTA (bottom panel) is able to resolve features, comparable to the gas map resolution (middle panel with no oversample), meaning the morphology can be further probed with CTA.

Our model provides a good framework for future studies, including insight into what parameters are required. A numerical approach in 3D could be used to more precisely model the particles injected, for each time step. This involves tracing the accelerated particles as they propagate and subsequently interact with the ISM, based on the magnetic field and diffusion parameters of each grid point in the model, hence it is a very computationally expensive approach.

For our model, we currently consider an impulsive injection of particles from the accelerator for the hadronic scenario. Mature aged SNRs may also produce \grays leptonically through the inverse-Compton effect \citep[e.g.][]{SN1006_IC, SN_Lep, RX_Lep}.
To finally confirm the accelerator of \hessj, we need to explore the leptonic scenario, assuming the SNR is accelerating electrons which may contribute to the TeV \gray emission as discussed by \cite{G8_Ajello, W30_liu_2019}. 
TeV emission from a PWNe is a reasonable assumption due to the turn-over in the GeV/TeV spectrum, which could be caused by cooling effects of electrons. Along with this, a continuous acceleration scenario should be investigated for both leptonic (typical of PWNe) and hadronic origins. Our model predicts an energy-dependent morphology, largely due to the bubble component, which is not implied by the GeV/TeV observations. Future work will include an energy-dependent morphology study of the H.E.S.S. data, to investigate this further. 
A dedicated study of the \textit{Fermi}-LAT data could also help to resolve the morphology of \grays at lower energies, in addition to the use of CTA in the future.
These methods will hopefully help reveal the nature of \hessj.

\section*{Acknowledgements}
The Mopra telescope is part of the ATNF which is funded by the Australian Government for operation as a National Facility managed by CSIRO (Commonwealth Scientific and Industrial Research Organisation). Support for observations were provided by the University of New South Wales and the University of Adelaide. K.F. acknowledges support through the provision of Australian Government Research Training Program Scholarship. 
\section*{Data Availability}
The data underlying this article are available on the MopraGam website at \url{http://www.physics.adelaide.edu.au/astrophysics/MopraGam/}. Other datasets were derived from sources in the public domain: HGPS at \url{https://www.mpi-hd.mpg.de/hfm/HESS/hgps/} and SGPS at \url{https://www.atnf.csiro.au/research/HI/sgps/fits_files.html}.

\bibliographystyle{mnras}
\bibliography{mnras_template}

\appendix

\section{Position-velocity plots}
\label{apn:PV_plot}
\begin{figure}
\includegraphics[width=\columnwidth]{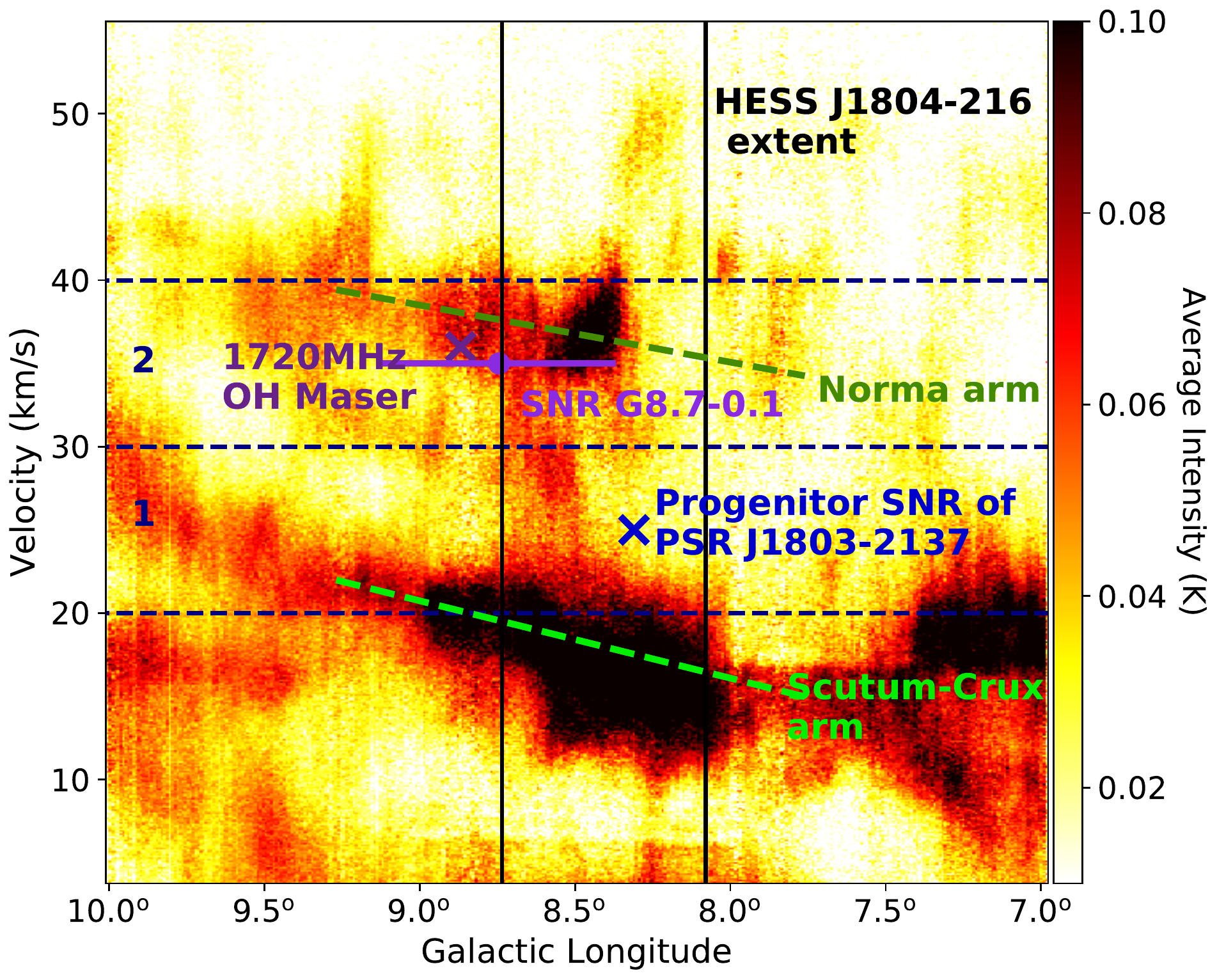}
\caption{Position-velocity plot of Mopra $^{12}$CO(1-0) emission (K) towards \hessj, integrated over latitudes from $-0.49^{\circ}$ to $0.49^{\circ}$ . The black vertical lines show the longitudinal extent of \hessj. The blue cross indicates the birth position of \psrj at its assumed velocity of ${\sim}25$\,\kms. The 1720MHz OH maser is shown by the purple cross at its velocity of 36\,\kms. The centre of \snrg is shown by the purple dot, whilst the purple line shows its radial extent, at a velocity of 35\,\kms. The green dashed lines are estimates of the Galactic spiral arms along the line of sight for \hessj \citep[from the model in][]{GRC_Vallee_2014}. Components 1 and 2 are indicated by the dashed navy lines.}
\label{fig:PVplot_CO}
\end{figure}

\begin{figure}
\includegraphics[width=\columnwidth]{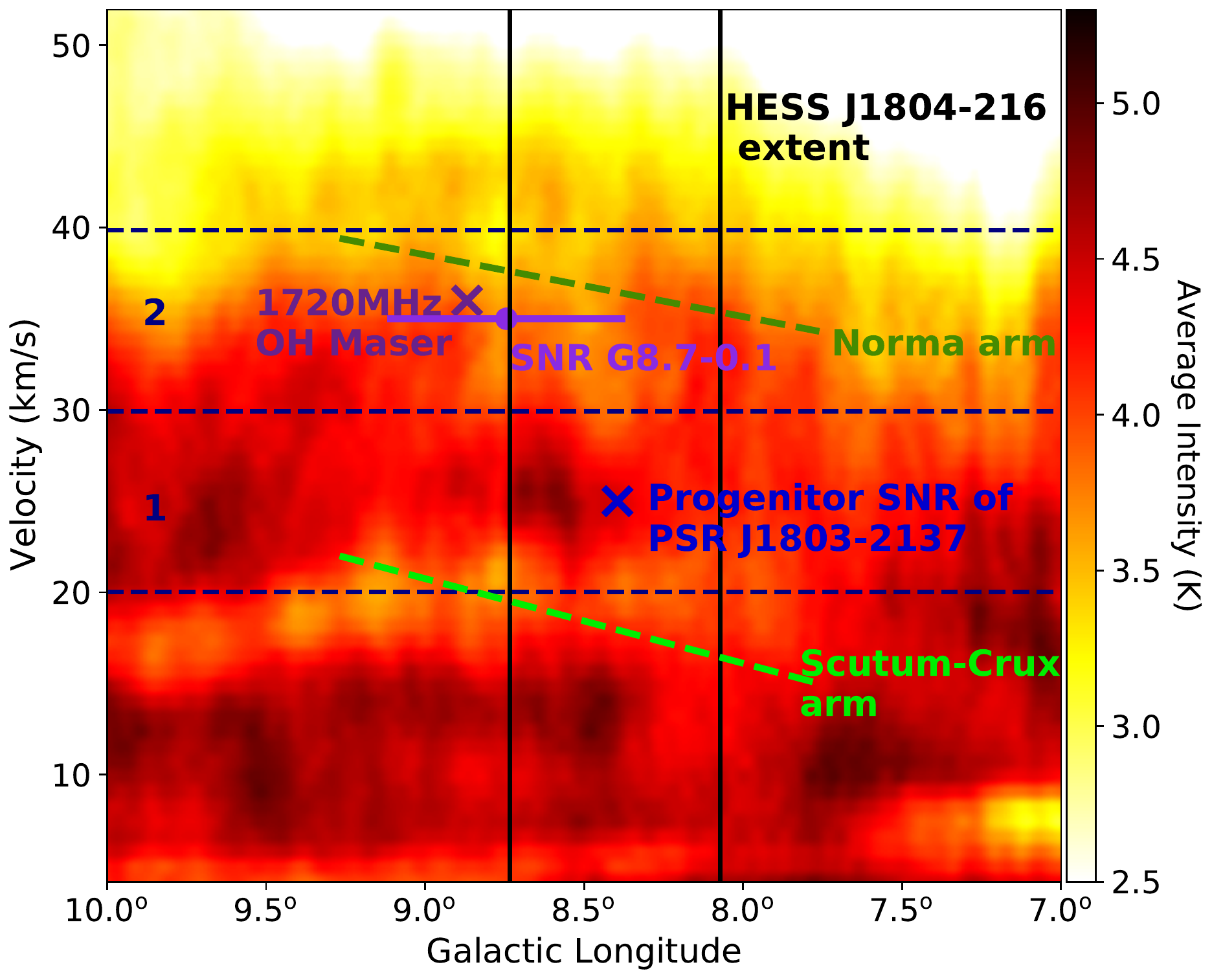}
\caption{Position-velocity plot of SGPS HI emission (K) towards \hessj. The annotations are described in \autoref{fig:PVplot_CO}.}
\label{fig:PVplot_HI}
\end{figure}

\cref{fig:PVplot_CO,fig:PVplot_HI} are position-velocity plots for the $^{12}$CO(1-0) and HI data, respectively, towards the \hessj region, integrated over latitudes $b=-0.49^{\circ}\,\rm{to}\,0.49^{\circ}$. Using this figure we defined two gas velocity regions, Component 1 (\vlsr=\ 20\ to\ 30\,\kms) and Component 2 (\vlsr=\ 30\ to\ 40\,\kms), as shown by the dashed navy lines. 

\section{Additional model terms}
\label{apn:params}
The onset of the Sedov-Taylor phase is defined in \citet{SNRs_Celli_2019}:

\begin{equation}
\tsedov \sim 1.6\times 10^3   \left( \dfrac{\ESN}{10^{51} \rm{erg}} \right) ^{-1/2} \left( \dfrac{\Mej}{10 M_\odot} \right)^{5/6} \rm{yr}
\label{eqn:t_sedov}
\end{equation}

where $\ESN$ is the ejected supernova total kinetic energy and $\Mej$ is the mass of the ejecta.

The inelastic cross-section of proton-proton collisions is taken from the most recent parameterisation by \citet{2014_Kafexhiu}.

\begin{equation}
\begin{split}
\sigma_{\rm{pp}}(\Ep) = & \left[ 30.7 - 0.96 \ln \left(\dfrac{\Ep}{\Ep^{\rm{th}}} \right) + 0.18 \ln\left( \dfrac{\Ep}{\Ep^{\rm{th}}}\right)^2 \right] \\
& \times \left[ 1 - \left( \dfrac{\Ep^{\rm{th}}}{\Ep} \right)^{1.9} \right]^3 \qquad \rm{mbarn}
\end{split}
\label{eqn:xsec_Kaf}
\end{equation}

where $\Ep$ is the kinetic energy of the proton and $\Ep^{\rm{th}}=2m_\pi + m_\pi^2/2m_{\rm{p}} \sim 0.2797$\,GeV is the threshold kinetic energy.

The total $\gamma$-ray spectrum is given by \cref{eqn:gray_spec}, for simplicity $x=E_\gamma/\Ep$.

\begin{multline}
  F_\gamma(x, \Ep) =
  B_\gamma \dfrac{\ln(x)}{x}
  \left( \dfrac{1-\xb }{1+\kg\xb(1-\xb)} \right)^4 \times  \\
  \left[ \dfrac{1}{\ln(x)} - \dfrac{4\bg\xb}{1-\xb}
  - \dfrac{4\kg\bg\xb(1-2\xb)}{1+\kg\xb(1-\xb)} \right]
\label{eqn:gray_spec}
\end{multline}

The additional parameters, $B_\gamma$, $\bg$ and $\kg$, are an approximation from numerical calculation using the best least square fit and dependent on the energy of the CR protons. For the proton energy range from $0.1-10^5$\,TeV these parameters are defined as:
\setlength{\abovedisplayskip}{5pt} \setlength{\abovedisplayshortskip}{5pt}
\begin{eqnarray}
B_\gamma &= & 1.3+0.14L+0.011L^2 \nonumber \\
\bg &= & (1.79+0.11L+0.008L^2)^{-1}   \\
\kg &= & (0.801+0.049L+0.014L^2)^{-1} \nonumber
\end{eqnarray}

\section{\hessj spectral comparison}
\begin{figure}
\begin{center}
\includegraphics[width=\columnwidth]{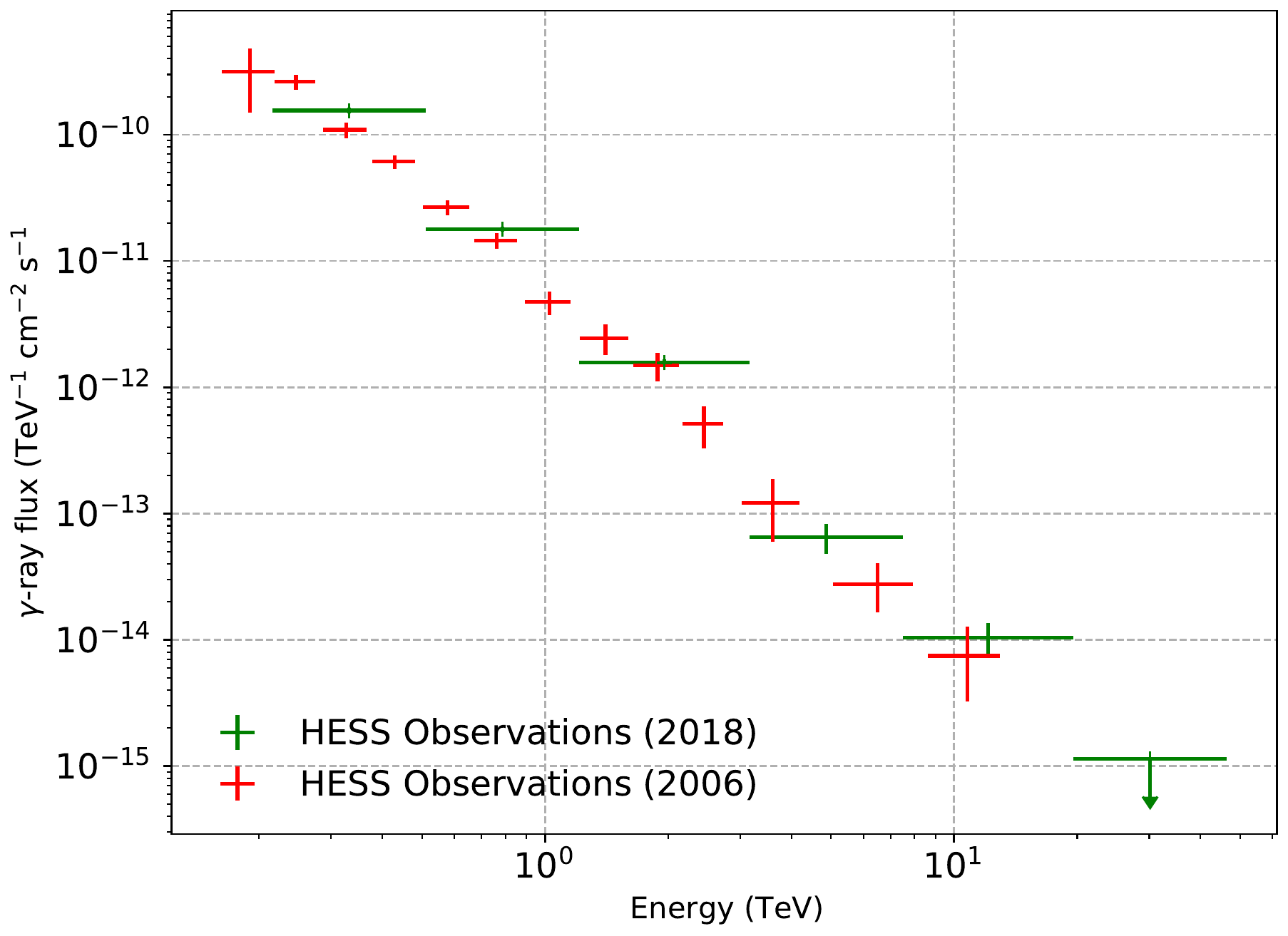}
\caption{TeV \gray spectra towards \hessj from \citet{2006_HESS} in red and \citet{2018_HGPS} in green. \citet{2006_HESS} has more spectral points as this involved a dedicated source analysis on \hessj.}
\label{fig:2006_2018}
\end{center}
\end{figure}

\autoref{fig:2006_2018} shows the spectral comparison between the H.E.S.S. 2006 survey of the inner galaxy \citep{2006_HESS} and the 2018 HGPS data \citep{2018_HGPS} towards \hessj. These spectra tend to match well. The observations from \cite{2006_HESS} are between May and July of 2004 with 11.7\,hr of observation time. The data from \cite{2018_HGPS} was collected between January of 2006 and January of 2013 with ${\sim}$44\,hr of observations. \cite{2006_HESS} provide a dedicated source analysis on \hessj with more spectral data points than \cite{2018_HGPS} which has a total of six bins (fixed bin number for all sources). The data from \cite{2006_HESS} is utilised here, as we want to compare the spectral shape of the observations with our model.

\section{Best matching models}
The following tables show the model parameters for the best matching spectral models based on their $\chi^2/m$ values, shown here in ascending order. Each model is given a name identifier, used within the main text. 

\begin{table*}
\begin{center}
\caption{The 5 best matching spectral models for accelerator \snrg for an age of 15\,kyr with their model parameters.}
\begin{tabular}{ccccccccccccc}
\hline 
Model & $\alpha$ & $\chi$ & $\delta$ & $\deltap$  & $\Epmax$ & SN Type &  $\Mej$ & $\ESN$ &       $\Wp$       & $\Epesc$  & $\chi^2/m$ &    $S$   \\
name &           &        &          &             &       (PeV)      &         &  ($M_\odot$)   &      (erg)    & ($10^{49}$\,erg)  &       (TeV)       & (Spectral)     & (Spatial, $10^{-13}$) \\
\hline
G1a  & 2.0   &  0.001 &  0.5   &   1.4   &   5.0    &   Ia   & 1  & $10^{51}$ &  7.0  & 14.8  &   1.0   &  2.8  \\
G2a  & 2.0   &  0.001 &  0.4   &   2.5   &   5.0    &   II   & 10 & $10^{51}$ &  6.9  & 18.6  &   1.1   &  3.2  \\
G3a  & 2.0   &  0.01  &  0.3   &   1.4   &   5.0    &   Ia   & 1  & $10^{51}$ &  7.0  & 14.8  &   1.1   &  3.0  \\
G4a* & 2.0   &  0.01  &  0.6   &   2.5   &   1.0    &   II   & 20 & $10^{51}$ &  6.2  & 15.7  &   1.1   &  2.7  \\
G5a  & 2.0   &  0.001 &  0.4   &   2.5   &   1.0    &   II   & 20 & $10^{51}$ &  6.2  & 15.7  &   1.1   &  2.8  \\
\hline
\end{tabular}
\\ \small * Best spatial model
\label{tab:G8_15kyr}
\end{center}
\end{table*}

\begin{table*}
\begin{center}
\caption{The 5 best matching spectral models for accelerator \snrg for an age of 28\,kyr  with their model parameters.}
\begin{tabular}{ccccccccccccc}
\hline 
Model & $\alpha$ & $\chi$ & $\delta$ & $\deltap$  & $\Epmax$ & SN Type &  $\Mej$ & $\ESN$ &       $\Wp$       & $\Epesc$  & $\chi^2/m$ &    $S$   \\
name &           &        &          &             &       (PeV)      &         &  ($M_\odot$)   &      (erg)    & ($10^{50}$\,erg)  &       (TeV)       & (Spectral)     & (Spatial, $10^{-13}$) \\
\hline
G1b  & 1.8   &  0.01  &  0.6   &   1.4   &   5.0    &   Ia   & 1  & $10^{51}$ &  1.5  & 6.2  &   0.6   &  1.9  \\
G2b* & 1.8   &  0.1   &  0.4   &   1.4   &   5.0    &   Ia   & 1  & $10^{51}$ &  1.5  & 6.2  &   0.8   &  1.6  \\
G3b  & 1.8   &  0.001 &  0.4   &   1.4   &   1.0    &   II   & 20 & $10^{52}$ &  1.0  & 8.1  &   1.0   &  2.3  \\
G4b  & 2.0   &  0.001 &  0.4   &   1.4   &   5.0    &   II   & 10 & $10^{52}$ &  0.7  & 18.1 &   1.0   &  3.2  \\
G5b  & 1.8   &  0.1   &  0.4   &   1.4   &   1.0    &   II   & 20 & $10^{52}$ &  1.0  & 8.1  &   1.0   &  2.0  \\
\hline
\end{tabular}
\\ \small * Best spatial model
\label{tab:G8_28kyr}
\end{center}
\end{table*}

\begin{table*}
\begin{center}
\caption{The 5 best matching spectral models for accelerator progenitor SNR of \psrj for an age of 16\,kyr  with their model parameters.}
\begin{tabular}{cccccccccccccc}
\hline 
Model & $\alpha$ & $\chi$ & $\delta$ & $\deltap$ &  $n_0$      & $\Epmax$ & SN Type &  $\Mej$ & $\ESN$ &     $\Wp$     & $\Epesc$ & $\chi^2/m$ &    $S$   \\
name &           &        &          &            & (cm$^{-3}$) &       (PeV)      &         &  ($M_\odot$)   &      (erg)    & ($10^{49}$\,erg)  &   (TeV)     &   (Spectral)     & (Spatial, $10^{-13}$) \\
\hline
P1* & 1.8   &  0.01  &  0.6   &   2.5   &  20  &   5.0    &   II   & 20 & $10^{52}$ & 3.8   & 3.8   &   0.9   &  1.2  \\
P2  & 1.8   &  0.01  &  0.6   &   2.5   &  10  &   5.0    &   II   & 20 & $10^{52}$ & 4.5   & 3.8   &   0.9   &  1.4  \\
P3  & 1.8   &  0.001 &  0.3   &   1.4   &  20  &   1.0    &   II   & 10 & $10^{52}$ & 2.6   & 8.0   &   1.0   &  2.6  \\
P4  & 1.8   &  0.01  &  0.6   &   2.5   &  1   &   5.0    &   II   & 20 & $10^{52}$ & 8.7   & 3.8   &   1.0   &  2.6  \\
P5  & 2.0   &  0.001 &  0.4   &   2.5   &  0.1 &   1.0    &   II   & 20 & $10^{51}$ & 3.5   & 13.4  &   1.0   &  1.2  \\
\hline
\end{tabular}
\\ \small * Best spatial model
\label{tab:PJ_16kyr}
\end{center}
\end{table*}

\bsp
\label{lastpage}
\end{document}